\documentclass[epj]{svjour}
\usepackage{graphics}
\usepackage[normalem]{ulem} 
\renewcommand\sout{\bgroup \color{red} \ULdepth=-.5ex \ULset}
\usepackage[dvips]{color} 

\begin{document}
\title{Stable multiquark states with heavy quarks in a diquark model}
\author{Su Houng Lee\inst{1}\thanks{e-mail: suhoung@phya.yonsei.ac.kr}
 \and Shigehiro Yasui\inst{2}
%
\thanks{e-mail: yasuis@post.kek.jp}
}                     
%
%
\institute{Institute of Physics
and Applied Physics, Yonsei University, Seoul 120-749, Korea \and Institute of Particle and Nuclear Studies, High Energy Accelerator Research Organization (KEK), 1-1, Oho, Ibaraki 305-0801, Japan}
\date{Received: date / Revised version: date}
%
\abstract{
Based on the color-spin interaction in diquarks, we argue why some multiquark configurations could be stable against strong decay when heavy quarks are included.  After showing the stability of previously discussed states we identify new possible stable states.  These are the $T^0_{cb}(ud \bar{c}\bar{b})$  tetraquark, the $\Theta_{bs}(udus \bar{b})$ pentaquark and the $H_c(udusuc)$ dibaryon, and so forth.
} 
\maketitle
\section{Introduction}

Multiquark configuration for the scalar nonet was first suggested by Jaffe in the framework of the MIT
bag model \cite{Jaffe76}.  Recently, the discussion on possible multiquark mesons has gained more ground as the recently discovered Z(4430) \cite{Z4430}, Z(4051) and Z(4248) \cite{Z4051} seem to be composed of $c\bar{c} ud$ type of four quarks, as they decay into $\psi' \pi$ and $\chi_{c1}\pi$ respectively.  The discussion on possibly multiquark configurations that are stable against strong decay is an old one starting from the $H$ dibaryon ($udsuds$) \cite{Jaffe-H}, multiply charmed $H$ dibaryon ($udsccc$) or ($uudscc$) \cite{SchaffnerBielich:1998ci}, charmed pentaquarks $\Theta_{cs}$ ($udus\bar{c}$) \cite{Gignoux:1987cn,Lipkin87,Riska:1992qd,Oh:1994yv,Oh:1997tp,Stewart:2004pd,Sarac:2005fn} to tetraquarks $T_{cc}$ ($ud\bar{c}\bar{c}$) \cite{Carlson:1988hh,Brac93,Manohar:1992nd}.  While there are different models predicting the stability of these states, in the constituent quark model, the stability of possible stable multiquark configurations originate from the strong scalar diquark attraction \cite{Jaffe76,Jaffe03}.  The strong attraction in the color anti-triplet channel is an important ingredient in hadron spectroscopy and is also responsible for color superconductivity at high density \cite{Alford:1997zt}.  In that sense, finding a stable multiquark configuration is also an important step toward understanding QCD at high density \cite{Lee:2007tn}.

In this work, we introduce a simple model that can be used to analyze the stability of multiquark configuration against strong decay and show why some of  the previously discussed states could indeed be stable. Then we discuss several new stable multiquark states, namely tetraquark, pentaquark and dibaryon containing charm and bottom  quarks.  The paper is organized as follows.  In section II, we introduce our simple constituent quark model.  Based on this model, we discuss possible stable multiquark states and their measurable decay modes in section III.  The summary is given in section IV.

\section{A schematic model for hadron mass differences}


It is a well known fact that the color-spin interaction \cite{DeRujula75} plays an important role in hadron mass differences. To illustrate
the mechanism, we introduce the following simplified form for the
color-spin interaction :
\begin{eqnarray}
C_H \sum_{i>j} \vec{s}_i \cdot \vec{s}_{j} \frac{1}{m_i m_j}.
\label{mass1}
\end{eqnarray}
Here $m_i$ and $\vec{s}_i$ are the mass and spin of the constituent
quark $i$. The overall strength of the color-spin interaction is given as \cite{KerenZur:2007vp}
\begin{eqnarray}
C_H=v_0 \vec{\lambda}_i \cdot \vec{\lambda}_j \langle \delta(r_{ij}) \rangle,
\label{coupling}
\end{eqnarray}
where $v_0$ is related to the coupling, the second part is the color factor and the third part is the strength of the wave function at zero separation.   The color factor $\vec{\lambda}_i \cdot \vec{\lambda}_j$ would be $-8/3$ for
diquarks in the color antitriplet channel and $-16/3$ for quark and
anti-quark pair in the color singlet channel.   It is a well known fact that the wave function effect can almost be treated as a constant \cite{Lee:2007tn,Rosner:2006yk}. To illustrate this point, we assume the following constituent quark
masses: $m_{u,d}=300~{\rm MeV}$, $m_s=500~{\rm MeV}$, $m_c=1500~{\rm
MeV}$, and $m_b=4700~{\rm MeV}$.

\begin{table}[h]
\centering
\caption{Baryon mass relations.  The first column is fit to
experiments.} \label{baryon}
\begin{tabular}{|c|c|c|c|c|}
\hline  Mass & $M_\Delta\!\!-\!\!M_N$  & $M_\Sigma\!\!-\!\!M_\Lambda$
& $M_{\Sigma_c}\!\!-\!\!M_{\Lambda_c}$ & $M_{\Sigma_b}\!\!-\!\!M_{\Lambda_b}$
 \\
 [2pt]
  \hline Form. & $\frac{3C_B}{2 m_u^2}$
& $\frac{C_B}{m_u^2}(1\!\!-\!\! \frac{m_u}{m_s})$ &
$\frac{C_B}{m_u^2}(1\!\!-\!\! \frac{m_u}{m_c})$
&
$\frac{C_B}{m_u^2}(1\!\!-\!\! \frac{m_u}{m_b})$
 \\
 [2pt]
Fit   & 290 MeV   &  77 MeV  & 154 MeV  & 180 MeV    \\[2pt]
Exp. & 290 MeV & 75 MeV & 170 MeV & 192 MeV \\[2pt] \hline
\end{tabular}
\end{table}

Table \ref{baryon} shows the mass differences between baryons that
are sensitive to the color-spin interaction only. By fitting $C_B$
to $M_\Delta - M_N$, we obtain ${C_B}/{m_u^2}=193$ MeV and find that
the mass differences $M_\Sigma-M_\Lambda$,
$M_{\Sigma_c}-M_{\Lambda_c}$ and $M_{\Sigma_b}-M_{\Lambda_b}$ are well reproduced. This is in no way
an attempt to make the best fit, but the point is that with typically
accepted constituent quark masses, the mass splittings that are
sensitive to the color-spin interaction are well reproduced with a
single parameter. As can be seen from Table~\ref{meson}, the fit is
equally good for the meson masses with ${C_M}/{m_u^2}=635$ MeV.
Here, even the $B^*-B$ mass difference fits well within the scheme.
From these fits, we find that $C_M$ is about 3 times larger than
$C_B$, reflecting that the quark and anti-quark correlation is about
3 times stronger than that between two quarks.
The reason why the wave function effect can be treated as a constant is a dynamical question, and some insight can be obtained by looking at various potentials \cite{KerenZur:2007vp}.   Within a naive bag model description, the wave function at the origin is only determined by the size of the Bag.  Hence, assuming a constant bag radius one finds a constant strength of the wave function at the origin.

\begin{table}[h]
\centering
\caption{Meson mass relations.  The first column is fit to
experiments.} \label{meson}
\begin{tabular}{|c|c|c|c|c|}
\hline  Mass & $M_\rho\!-\!M_\pi$ &
$M_{K^*}\!-\!M_{K}$  & $M_{D^*}\!-\!M_D$  &  $M_{B^*}\!-\!M_B$ \\[2pt] \hline
Form. & $\frac{C_M}{m_u^2}$ &  $\frac{C_M}{m_u m_s}$  &
$\frac{C_M}{m_u m_c}$  &   $\frac{C_M}{m_u m_b}$  \\[2pt]
Fit & 635 MeV   &  381 MeV  & 127 MeV    &  41 MeV \\[2pt]
Exp. & 635 MeV &  397 MeV & 137 MeV & 46 MeV  \\[2pt] \hline
\end{tabular}
\end{table}

When both quarks are heavy, the value of  $C_H$ is expected to
become larger as the strength of the relative wave function at the
origin is substantially increased. Fitting instead its value to the
mass difference between $J/\psi$ and $\eta_c$, we find
$C_{c\bar{c}}/{m_c^2}=117$ MeV. Assuming that the corresponding
attraction between charmed diquark is three times smaller than that
between the charm quark-antiquark pair as in the case of light
quarks, we have $C_{cc}/{m_c^2}=39$ MeV.
In the mass difference between $\Upsilon$ and $\eta_{b}$, in which the latter has been recently observed by BaBar \cite{:2008vj}, we also find $C_{b\bar{b}}/{m_b^2}=71.4$ MeV, and in turn $C_{bb}/{m_b^2}=23.8$ MeV.
Therefore, we find $C_{bb}$ is much larger than $C_{cc}$.
Concerning the value of $C_{bc}$, there is no available experimental data for the mass difference between $B_{c}^{\ast}$ and $B_{c}$.
However, we also expect that $C_{cb}$ has flavor dependence and given by an value  interpolating between that for the $cc$ and the $bb$ diquarks.
In any case, we find the flavor dependence induces the larger value of $C_{H}$ for heavier flavors.
Therefore, we could introduce
additional mass dependence in $C_B$ and in $C_M$ by fitting the mass
differences in the strange, charm and bottom hadrons
from Table I and II, respectively.
However, these introduce only minor changes in the analysis to
follow, and therefore we will just use the mass independent $C_H$'s
obtained above.

\section{Stable multiquark configuration }

Using above parameters, one notes that the scalar diquark has a large attractive contribution to the mass.  Any multiquark configuration that are stable should be the ones that  are composed of such configurations.

\begin{table}[h]
\begin{center}
\caption{\small \baselineskip=0.5cm Type of attractive scalar diquarks $(q_{1}q_{2})$ composed of quark $q_1$ and $q_2$.  The second line is their binding in MeV due to the color-spin force $ - \frac{3}{4} \frac{C_{B}}{m_{q_1}m_{q_2}}$ .}
\label{tbl:diquarks}
\begin{tabular}{c|c|c|c}
\hline \hline
 $(ud)$ &  &  &  \\
 -144.75  &  &  & \\
\hline
 $(us)$ & $(ds)$ &  &  \\
 -86.85  & -86.85 & &  \\
\hline
 $(uc)$ & $(dc)$ & $(sc)$ &  \\
 -28.95 & -28.95 & -17.37 & \\
\hline
 $(ub)$  & $ (db)$ & $(sb)$ & $(cb)$ \\
  -9.23 & - 9.23 & -5.54 & -1.84  \\
\hline
\hline
\end{tabular}
\label{diquarks}
\end{center}
\end{table}%

Table \ref{tbl:diquarks} lists the possible diquarks and their attractions due to Eq.~(\ref{mass1}) with parameters obtained above.  Candidates for stable multiquark configurations can be obtained by combining diquarks and possible color source to make them color singlet.

\subsection{Heavy tetraquark}

Tetraquarks are obtained by combining a diquark and an antidiquark in Table \ref{tbl:diquarks}.   Let us start by looking at the tetraquark with total spin zero containing the $(ud)$ diquark and an anti-diquark of $(\bar{q}_1\bar{q}_2)$; the configuration will be called $T^0_{q_{1}q_{2}}$.   In the following discussions, we will only consider quarks in the $s$-state, and assume that the color-spin effect is as given in Eq.~(\ref{mass1}) with constant coefficients determined above.
To investigate the stability of the $T^0_{q_1q_2}$ against strong decay, one has to consider its
mass difference with the two pseudo scalar mesons $M_1(u\bar{q}_1)$ and $M_2(d\bar{q}_2)$, which we call the binding energy $B_{T^0_{q_{1}q_{2}}}$,
\begin{eqnarray}
B_{T^0_{q_{1}q_{2}}} & = & m_{T^0_{q_{1}q_{2}}}-m_{M_1}-m_{M_2} \nonumber \\
& = &  - \frac{3}{4} C_{B} \bigg( \frac{1}{m_u^2}+ \frac{1}{m_{q_1}m_{q_2}} \bigg)
\nonumber \\ && +\frac{3}{4} C_{M} \bigg( \frac{1}{m_u m_{q_1}}+ \frac{1}{m_{u}m_{q_2}} \bigg).
\label{mass-diff1}
\end{eqnarray}
Since phenomenologically $C_M$ is about 3 times larger than $C_B$, the combinations of diquarks which make Eq.~(\ref{mass-diff1})  negative are very limited.  Only when  $q_1=c$ and $q_2=b$, the binding energy becomes $B_{T^0_{cb}}=-21.25$ MeV, and the configuration $(ud\bar{c}\bar{b})$ which we call $T^0_{cb}$ will be stable against strong decay.  We note that more realistically the binding energy of a diquark with two heavy quarks could be larger as noted previously. Therefore the binding energy we have obtained will be the lower bound.

This result can be understood simply in the heavy quark limit.
The binding energy of $T^0_{cb}$ is supplied mainly from the binding energy of  $(ud)$ diquark, because the binding energy of $(cb)$ diquark is suppressed by $1/m_{c}m_{b}$.
On the other hand, the binding energies of the two meson states, $u\bar{c}$ and $d\bar{b}$, are also suppressed by $1/m_{c}$ and $1/m_{b}$, respectively.
Therefore, the $(ud)$ diquark is regarded as an attractive force which stabilizes $T^0_{cb}$.
Then, we may ask if $T^0_{cb}$ including $(us)$ or $(ds)$ diquarks can be stable, because $s$ quark may be considered to belong to the light quarks.
We find that the configuration $us\bar{c}\bar{b}$ could be stable against $B^{+}+D_{s}^{-}$ ($B_{T^0_{cb}(us)}=-1.1$ MeV), however it is unstable against $\bar{D}^{0}+B_{s}^{0}$ ($B_{T^0_{cb}(us)}=+24.8$ MeV) because of the relatively weaker binding in the $(us)$ diquark.
Therefore, $T_{cb}^{0}$ will not form $\bar{\bf 3}_{f}$ multiplet in light flavor $SU(3)_{f}$ symmetry.

Another tetraquark that could be stable is $T^1_{cc}$ that was discussed before within the present model in Ref.~\cite{Lee:2007tn}.  $T^1_{cc}$ is a state with scalar diquark $(ud)$ and axial vector diquark $(cc)$, hence the total quantum number is $J^P=1^+$.   This state can not decay strongly to two pseudo scalar mesons, but can to a pseudo scalar meson and a vector meson.  Therefore, although the charmed anti-diquark is repulsive, the attractive $(ud)$ diquark gives more binding than the pseudo scalar and vector mesons.  That is,
\begin{eqnarray}
B_{T^1_{cc}} & = & m_{T^1_{cc}}-m_{D}-m_{D^*} \nonumber \\
& = &  - \frac{3}{4}  \frac{C_{B}}{m_u^2}+ \frac{1}{4} \frac{C_{B}}{m_{c}^2}
 +\frac{3}{4} C_{M} \bigg( \frac{1}{m_u m_{c}}- \frac{1}{3m_{u}m_{c}} \bigg)
 \nonumber \\ &\sim& -79.3 ~{\rm MeV}.
\label{mass-diff2}
\end{eqnarray}
In the same way, $T_{bb}^{1}$ can also be a stable tetraquark, whose binding energy is given by
\begin{eqnarray}
B_{T^1_{bb}} & = & m_{T^1_{bb}}-m_{B}-m_{B^*} \nonumber \\
& = &  - \frac{3}{4}  \frac{C_{B}}{m_u^2}+ \frac{1}{4} \frac{C_{B}}{m_{b}^2}
+\frac{3}{4} C_{M} \bigg( \frac{1}{m_u m_{b}}- \frac{1}{3m_{u}m_{b}} \bigg)
\nonumber \\ &\sim& -124.3 ~{\rm MeV}.
\label{mass-diff2b}
\end{eqnarray}
We note that $T_{cc(bb)}^{1}$ containing $(us)$ or $(ds)$ diquark instead of $(ud)$ diquark can also be stable as shown in Table \ref{tetraquark1_mass}.
For example, $T_{cc}^{1}(us)$ has binding energy -8.7 MeV against $\bar{D}^{0}+D_{s}^{*-}$, and $T_{bb}^{1}(us)$ has -62.3 MeV against $B^{+}+B_{s}^{*0}$.
Therefore, $T_{cc(bb)}^{1}$ will form  $\bar{\bf 3}_{f}$ multiplet.

In between $T^{1}_{cc}$ and $T^{1}_{bb}$, $T^{1}_{cb}$ with spin triplet $(cb)$ diquark is also interesting.
The binding energy is obtained similarly as shown in Table \ref{tetraquark1_mass}.
In this case, $T^{1}_{cb}$ with $(ud)$ diquark is stable, while $T^{1}_{cb}$ with $(us)$ and $(ds)$ diquarks are not.
Therefore, $T^{1}_{cb}$ would not form $\bar{\bf 3}_{f}$ multiplet, as only $T^{1}_{cb}$ is bound.
$T^{1}_{cb}$ is regarded as an excited state of $T^{0}_{cb}$, whose  mass splitting is given by $C_{B}/m_{c}m_{b}$.
Therefore, if  both $T^{0}_{cb}$ and $T^{1}_{cb}$ are observed and their mass splitting fits well to $C_{B}/m_{c}m_{b}$, then it confirms that $T_{cb}$ is really a tetraquark, and not a bound molecular state, which could otherwise be another possibility.

\begin{table}[htdp]
\caption{
The binding energy $B_{T_{cc(bb,cb)}^{1}} = m_{T_{cc(bb,cb)}^{1}} - m_{M} - m_{M^*}$ of $T_{cc(bb,cb)}^{1}$ against decay to pseudoscalar and vector mesons, $M$ and $M^{*}$, as shown below. The unit is in MeV.
The tetraquarks can be stable for $B_{T_{cc(bb,cb)}^{1}}<0$.
}
\begin{center}
\begin{tabular}{c|c|c|c}
\hline \hline
& $ud\bar{c}\bar{c}$ & $us\bar{c}\bar{c}$ & $ds\bar{c}\bar{c}$ \\
\cline{2-4}
$T_{cc}^{1}$ & -79.3 & -8.7 & -8.7 \\
 & $\bar{D}^{0}+D^{*-}$, $\bar{D}^{* 0}+D^{-}$ & $\bar{D}^{0}+D_{s}^{*-}$ & $D^{-}+D_{s}^{*-}$ \\
\hline \hline
& $ud\bar{b}\bar{b}$ & $us\bar{b}\bar{b}$ & $ds\bar{b}\bar{b}$ \\
\cline{2-4}
$T_{bb}^{1}$ & -124.3 & -62.3 & -62.3 \\
 & $B^{+}+B^{*0}$, $B^{* +}+B^{0}$ & $B^{+}+B_{s}^{*0}$ & $B^{0}+B_{s}^{* 0}$ \\
\hline \hline
& $ud\bar{c}\bar{b}$ & $us\bar{c}\bar{b}$ & $ds\bar{c}\bar{b}$ \\
\cline{2-4}
$T_{cb}^{1}$ & -59.0 & +2.9 & +2.9 \\
 & $B^{*+}+D^{-}$, $B^{*0}+\bar{D}^{0}$ & $B_{s}^{*0}+D^{0}$ & $B_{s}^{*0}+D^{-}$ \\
\hline \hline
\end{tabular}
\end{center}
\label{tetraquark1_mass}
\end{table}%

The presence of $T_{cb}^{0}$ and $T_{cc}^{1}$ can be observed experimentally through their hadronic weak decay \cite{Lee:2007tn}.
$T_{cb}^{0}$ contains the components of $\bar{D}^{0}+B^{0}$ or $D^{-}+B^{+}$, whose available decay modes are listed in Ref.~\cite{Amsler:2008zz}.
For example, through the weak decay $\bar{D}^{0} \rightarrow K^{+}\pi^{-}$ and $B^{0}\rightarrow K^{+}\pi^{-}$, we could observe $T_{cb}^{0}\rightarrow K^{+}\pi^{-}+K^{+}\pi^{-}$, as far as $T_{cb}^{0}$ is really stable against the strong decay.  While the expected branching ratio for the particular hadronic decay mode is very small because of the small branching ratio of  $B^0\rightarrow K^{+}\pi^{-}$, we need to search for such modes to reconstruct the mass of $T_{cb}^0$.
Similarly, $T_{cc}^{1}$ contains $\bar{D}^{0}+D^{* -}$ or $\bar{D}^{* 0}+ D^{-}$.
Then, using $D^{*} \rightarrow D\pi$, an interesting decay mode is given as $T_{cc}^{1} \rightarrow K^{+}\pi^{-}+K^{+}\pi^{-}+\pi^{-}$ \cite{Lee:2007tn}.
In the same way, $T_{bb}^{1}$ and $T^{1}_{cb}$ contain $B+B^{*}$ and $D+B^{*}$, respectively.
However, the prediction of decay modes are not easy at present because the available decay modes of $B^{*}$ are very limited.

\subsection{Heavy pentaquark}

Pentaquarks are obtained by taking any two diquarks as given in Table \ref{diquarks} together with an antiquark.   Stability in this case is guaranteed when the binding energy of the pentaquark configuration is larger than the largest sum of baryon and meson binding energies.  The diquarks should be of different types as they have to be in color antisymmetric configuration so that it combines with the antiquark into a color singlet baryon.   For the pentaquark to be stable, it would be the best to start with the diquarks with the largest bindings; that is $(ud)$ and $(us)$.  Combining them with an antiquark $q$, the binding energy of the pentaquark $\Theta_{qs}$ ($udus\bar{q}$) becomes,
\begin{eqnarray}
B_{\Theta_{qs}} & = & m_{\Theta_{qs}}-m_{\Lambda}-m_{u\bar{q}} \nonumber \\
& = &  - \frac{3}{4}  \frac{C_{B}}{m_u^2}- \frac{3}{4}  \frac{C_{B}}{m_u m_s}
\nonumber \\ && +\frac{3}{4} \frac{C_{B}}{m_u^2} + \frac{3}{4} \frac{C_M}{m_u m_q}. \label{mass-diff3}
\end{eqnarray}
The first and third term cancels.  Since $C_M$ is about larger than $C_B$ by a factor 3, it is clear that we will have binding only if $m_q > \frac{C_M}{C_B} m_s > 3 m_s$.  Therefore, the binding is slightly positive for $q=c$ with 8.4 MeV as shown in Table \ref{pentaquark_mass}.
For $q=b$, one can certainly expect a bound $\Theta_{bs}$ whose binding is $-56.4$ MeV \cite{Lee:2007tn}.
The large attraction gained from recombining it into a meson and a baryon is the reason why the light pentaquark can not be stable, as was shown in a more detailed constituent quark model \cite{Hiyama06}.

Next, we consider the configuration of $(us)$ and $(ds)$ diquarks with anti-quark $\bar{q}$, which we denote as $\Theta_{qss}(usds\bar{q})$.
In this case we have two possible final states; $\Xi+u\bar{q}$ and $\Lambda+s\bar{q}$.
The binding energy for the former state is given by
\begin{eqnarray}
B_{\Theta_{qss}} & = & m_{\Theta_{qss}}-m_{\Xi}-m_{u\bar{q}} \nonumber \\
& = &  -2 \times \frac{3}{4}  \frac{C_{B}}{m_u m_s} +\frac{3}{4} \frac{C_{B}}{m_{u} m_{s}} + \frac{3}{4} \frac{C_M}{m_u m_q},
\label{mass-diff4}
\end{eqnarray}
and that for the latter state by
\begin{eqnarray}
B_{\Theta_{qss}} & = & m_{\Theta_{qss}}-m_{\Lambda}-m_{s\bar{q}} \nonumber \\
& = &  -2 \times \frac{3}{4}  \frac{C_{B}}{m_u m_s} +\frac{3}{4} \frac{C_{B}}{m_{u}^{2}} + \frac{3}{4} \frac{C_M}{m_s m_q}.
\label{mass-diff5}
\end{eqnarray}
The numerical result is shown in Table \ref{pentaquark_mass}.
In these configurations, we again find that the pentaquark becomes more stable as the included antiquark becomes heavier.
Therefore, $\Theta_{cs}$ and $\Theta_{css}$ are slightly unstable, while  $\Theta_{bs}$ and $\Theta_{bss}$ are stable.  Nonetheless, it is worth searching for all these states in the experiments.

The flavor structure of the stable pentaquark is obtained by following light flavor $SU(3)_f$ symmetry.
In the present discussion, the stable pentaquark contains $(ud)$, $(us)$ and $(ds)$ diquarks together with a heavy quark, say $q=b$.
Because these diquarks belong to ${\bf \bar{3}}_f$ representation in $SU(3)_f$, the flavor decomposition is obtained as
\begin{eqnarray*}
{\bf \bar{3}}_{f} \times {\bf \bar{3}}_{f} = {\bf 3}^{a}_{f} + {\bf \bar{6}}^{s}_{f}.
\end{eqnarray*}
In our diquark model, ${\bf \bar{6}}^{s}_{f}$ vanishes because it is symmetric under the exchange of flavors and hence can be canceled by anti-symmetry in color.
Therefore, the pentaquarks obtained above belong to the  ${\bf 3}^{a}_{f}$ representation.

\begin{table}[htdp]
\caption{The binding energy of pentaquarks $B_{\Theta_{qs}} =m_{\Theta_{s\bar{q}}} - m_{M}-m_{B}$ ($B_{\Theta_{qss}} =m_{\Theta_{qss}} - m_{M}-m_{B}$) of $\Theta_{qs}(ud\, us\, \bar{q})$ ($\Theta_{qss}(us\, ds\, \bar{q})$) for $q=u$, $s$, $c$ and $b$, respectively, against decays to  meson $M$ and baryon $B$ states. The unit is in MeV.
The pentaquarks can be stable for $B_{\Theta_{qs}(\Theta_{qss})}<0$.}
\begin{center}
\begin{tabular}{c|c|c|c|c|c}
\hline
  $\Theta$ & $M\!+\!B$ & $\bar{u}$ & $\bar{s}$ & $\bar{c}$ & $\bar{b}$ \\
\hline \hline
 $ud\, us\, \bar{q}$ &  $ud \, s$+$u\bar{q}$ & 389.4 & 198.9 & 8.4 & -56.4 \\
\hline \hline
 $us\, ds\, \bar{q}$ & $ds \, s$+$u\bar{q}$ & 389.4 & 198.9 & 8.4 & -56.4 \\
\cline{2-6}
                              & $ud \, s$+$s\bar{q}$ & 256.8 & 142.5 & 28.4 & -10.7 \\
\hline
\end{tabular}
\end{center}
\label{pentaquark_mass}
\end{table}%

The pentaquarks have easily identifiable decay modes.
For example, $\Theta_{cs}$ ($udds\bar{c}) \rightarrow \Lambda + K^{+} \pi^{-}$ \cite{Lee:2007tn}, and $\Theta_{bs}$ ($udds\bar{b}$)  $\rightarrow \Lambda + K^{+}\pi^{-}$, as modes with the least number of particles in the final states.
Similarly, $\Theta_{css} \rightarrow \Lambda + \pi^{+} \pi^{-} \pi^{-}$, and $\Theta_{bss} \rightarrow \Lambda + \pi^{+} \pi^{-} \pi^{-} + \pi^{+}$.

\subsection{Heavy dibaryon}

Dibaryons are obtained by taking three diquarks  and combining them into a color singlet combination. Considering the color anti-triplet scalar diquarks $(q_{i}q_{j})$ in Table \ref{diquarks}, the wave function is given as,
\begin{eqnarray}
\epsilon_{abc} (q_1 q_2)^a(q_3 q_4)^b(q_5 q_6)^c,
\label{H-wave}
\end{eqnarray}
where $a,b,c$ are the color indexes.
The total number of states which can be obtained from these wave functions are 120.  However, only a few of them are stable against strong decay to two baryons.  To understand which configurations are stable, let us consider the decay of the dibaryon into two baryons in the following way. First of all, let us assume that the diquark $(q_{1}q_{2})$ is the most stable one among the three diquarks in  Eq.~(\ref{H-wave}).
If there are two most stable diquarks, one of them is assigned to $(q_{1}q_{2})$.
When the dibaryon decays to two baryons, there should be a process where some diquarks in the dibaryon are broken into quarks.
In the present discussion, because $(q_{1}q_{2})$ is the most stable diquark, the remaining two diquarks, $(q_{3}q_{4})$ and $(q_{5}q_{6})$, can be broken.
First possible process occurs when one diquark is broken.
When $(q_{5}q_{6})$ is more loosely bound than $(q_{3}q_{4})$, then the former diquark is broken to a pair of single quarks $q_{5}$ and $q_{6}$.
Then they are combined with $(q_{1}q_{2})$ and $(q_{3}q_{4})$ to form two baryons that can be represented as either $(q_{1}q_{2})q_{5}$ and $(q_{3}q_{4})q_{6}$, or $(q_{1}q_{2})q_{6}$ and $(q_{3}q_{4})q_{5}$. This process is shown by the following equation,
\begin{eqnarray}
{\rm conf}_1 & = &  (q_1 q_2) q_5 + (q_3 q_4) q_6, \nonumber \\
&& \mbox{or\hspace{0.5em}} (q_1 q_2) q_6 + (q_3 q_4) q_5.
\label{conf1}
\end{eqnarray}
Second possible process occurs when two diquarks are broken.
In this case, the diquarks $(q_{3}q_{4})$ and $(q_{5}q_{6})$ are first broken into $q_{3}$, $q_{4}$, $q_{5}$ and $q_{6}$.
Second, when the most stable diquark among possible combinations of these four quarks is $(q_{3}q_{5})$, then they are combined to two possible final baryon states that are either $(q_{1}q_{2})q_{4}$ and $(q_{3}q_{5})q_{6}$, or $(q_{1}q_{2})q_{6}$ and $(q_{3}q_{5})q_{4}$.
This process is shown by,
\begin{eqnarray}
{\rm conf}_2 & = &  (q_1 q_2) q_4 + (q_3 q_5) q_6, \nonumber \\
&& \mbox{or\hspace{0.5em}} (q_1 q_2) q_6 + (q_3 q_5) q_4.
\label{conf2}
\end{eqnarray}
The stability of dibaryons is determined by the mass difference between the dibaryon and two baryon states.
Namely, when the mass of dibaryon is smaller than those of ${\rm conf}_{1}$ and ${\rm conf}_{2}$, then such dibaryon can be stable against the strong decay.
Now we investigate the stability of the configurations of ${\rm conf}_1$ or ${\rm conf}_2$.

Let us first consider the stability of dibaryon against the configuration ${\rm conf}_{1}$.
In the present diquark model, the diquark $(q_{5}q_{6})$ is a bound state for any flavor combinations with finite binding energy as given in Table~\ref{diquarks}.
Therefore, the mass of dibaryon is always smaller than that of  ${\rm conf}_{1}$ by the binding energy of $(q_{5}q_{6})$. Consequently, we obtain the following rule.
\begin{itemize}
\item[i)] Any dibaryon constructed from Table~\ref{diquarks} is stable against the decay to ${\rm conf}_{1}$.
\end{itemize}
However,  the dibaryons are not necessarily stable against the decay to ${\rm conf}_2$.
To see this, let us analyze the stability against the decay to ${\rm conf}_{2}$ in Eq.~(\ref{conf2}).  For this purpose, we consider the difference between the binding energies of the dibaryon and ${\rm conf}_{2}$.
Omitting the common binding energy supplied from $(q_{1}q_{2})$, the binding energy of the dibaryon is given by
\begin{eqnarray}
{\rm B.E.}\bigg [(q_3q_4)+(q_5q_6) \bigg ] & = & -\frac{3}{4}\frac{C_B}{m_3m_5} \bigg(\frac{m_5}{m_4}+\frac{m_3}{m_6} \bigg), \label{be1}
\end{eqnarray}
while that of ${\rm conf}_{2}$ is given by
\begin{eqnarray}
{\rm B.E.}\bigg [(q_3q_5) \bigg ] & = & -\frac{3}{4}\frac{C_B}{m_3m_5}. \label{be2}
\end{eqnarray}
As can be seen above, the dibaryon is stable if the following condition is satisfied,
\begin{eqnarray}
\frac{m_5}{m_4}+\frac{m_3}{m_6}  > 1 .
\label{cond1}
\end{eqnarray}
From this equation, we  obtain two rules to realize a stable dibaryon.
\begin{itemize}
\item[ii-a)] First, we considering the case where the two diquarks are in the same row in Table~\ref{diquarks}; $q_4=q_6$. Then condition (\ref{cond1}) reduces to
\begin{eqnarray}
m_3+m_5 > m_4.
\label{cond2}
\end{eqnarray}
Because two diquarks are taken from a single row in Table~\ref{diquarks}, we have the following possibilities; $\{(q_{3}q_{4}), (q_{5}q_{6})\}=\{(us), (ds)\}$ for $q_{4}=s$, $\{(uc), (dc)\}$, $\{(uc), (sc)\}$ and $\{(dc), (sc)\}$ for $q_{4}=c$, $\{(ub), (db)\}$, $\{(ub), (sb)\}$, $\{(ub), (cb)\}$, $\{(db), (sb)\}$, $\{(db), (cb)\}$ and $\{(sb), (cb)\}$ for $q_{4}=b$. Note $q_{3} \neq q_{5}$ from Eq.~(\ref{conf2}).
Among them, condition (\ref{cond2}) is satisfied only for $\{(q_{3}q_{4}), (q_{5}q_{6})\}=\{(us), (ds)\}$.
In other words, diquarks which contain at least two diquarks from either the third or the fourth row of Table~\ref{diquarks}, such as $\{uc, dc\}$, can not be stable against ${\rm conf}_2$.
\item[ii-b)] Second, we consider the case where the two diquarks are in different row of Table~\ref{diquarks}; $q_{4} \neq q_{6}$.  In this case however, one finds that no configuration satisfies Eq.~(\ref{cond1}) when $q_3$ and $q_5$ are different light quarks.

\end{itemize}
Combining these two rules, one can obtain stable dibaryon configurations by starting from either the first or second row and picking up the remaining two diquarks by moving down in rows either straight along a single column or to different columns with Eq.~(\ref{cond1}) satisfied.
Following these rules about decays to ${\rm conf}_1$ and ${\rm conf}_2$, it is easy to find that there are 16 stable dibaryons containing $(ud)$ diquark and 9 more containing $(us)$ or $(ds)$ diquarks.  The only exception to this rule is the $(ud)(sc)(sb)$ as it can decay into $(us)c+(ds)b$ type baryons. Table \ref{dibaryon} lists all the possible configurations and the stable dibaryons  whose binding energy are negative.

In Table \ref{dibaryon}, the stable dibaryons are categorized according to their  contents of diquarks as represented in the first column; $(qq)(qq)(qq)$, $(qq)(qq)(qQ)$, $(qq)(qQ)(qQ')$, $(qq)(qq)(QQ')$ and $(qq)(qQ)(QQ')$ with $q=u$, $d$, $s$ and $Q$, $Q'=c$, $b$ which may be called $H$, $H_{Q}$, $H_{Q'Q}$, $H'_{Q'Q}$ and $H_{Q'QQ}$, respectively.
The second column indicates the flavor content through five integers  $N_{u}N_{d}N_{s}N_{c}N_{b}$ representing the number of quark with $u$, $d$, $s$, $c$ and $b$ flavors respectively.
The third column is the configurations of diquarks in dibaryons. Here, for short description, we use $qq$ instead of $(qq)$.
The fourth, fifth and sixth columns are isospin, its $z$ component and strangeness, respectively.
The seventh column indicates the configurations of the  two baryon states with lowest threshold. The binding energy of the dibaryon against the two baryon threshold is shown as their mass difference in the last column.

The $(qq)(qq)(qq)$ type becomes stable from the rule ii-a) for ${\rm conf}_{2}$,
while the other configurations are stable from the rule i) for ${\rm conf}_{1}$.
We mention that several dibaryons which are not accounted from the rules i), ii-a) and ii-b) are also stable as components in isospin multiplets as shown in Table~\ref{dibaryon}.
They are the flavor $N_{u}N_{d}N_{s}N_{c}N_{b}=22110$ with $I=1$ and $I_{z}=0$ in $(qq)(qq)(qQ)$ ($Q=c$), 22101 with $I=1$ and $I_{z}=0$ in $(qq)(qq)(qQ)$ ($Q=b$), 21111 and 12111 with $I=3/2$ and $I_{z}=\pm1/2$ in $(qq)(qQ)(qQ)$, and 11121 with $I=1$, $I_{z}=0$ in $(qq)(qQ)(QQ')$.

Among these dibaryons, the following two dibaryon configurations are the most deeply bound states and are worth mentioning in details.  The first one is the $H$ dibaryon \cite{Jaffe-H} with diquark configurations of $(ud)(us)(ds)$.
Again, the binding energy of the $H$ dibaryon can be obtained by comparing its  color-spin interaction to that of two $\Lambda$'s,
\begin{eqnarray}
B_H & = & m_{H}-m_{\Lambda}-m_\Lambda \nonumber \\
& = &  - \frac{3}{4}  \frac{C_{B}}{m_u^2}- 2 \frac{3}{4}  \frac{C_{B}}{m_u m_s}
\nonumber \\ && +\frac{3}{4} \frac{C_{B}}{m_u^2} + \frac{3}{4} \frac{C_B}{m_u^{2} }. \label{mass-diff3}
\end{eqnarray}
The first term cancels the third term.  Therefore, the stability of $H$ dibaryon depends on whether it is energetically favorable to have $(us)+(ds)$ diquarks for a strong $(ud)$ diquark, which in fact is the rule {ii-a)} above.  Substituting the constants in our model, we find a binding of $B_H=-29$ MeV, which is consistent with most of the model calculations \cite{Sakai:1999qm}.

The second one is the dibaryon of the $(qq)(qq)(qQ)$ type with $Q=c$ containing a single charm, which we will call the $H_{c}$ dibaryon.
The most attractive dibaryon in such configuration is $(ud)(us)(uc)$,   $(ud)((us)(dc)+(ds)(uc))$ and $(ud)(ds)(dc)$ in the isospin $I=1$ multiplet.
Concerning $(ud)(us)(uc)$, the binding energy of $H_c$ can be investigated by comparing its color-spin interaction to that of the proton $(udu)$ and $\Xi_c$ $(usc)$,
\begin{eqnarray}
B_{H_c} & = & m_{H_c}-m_{p}-m_{\Xi_c} \nonumber \\
& = &  - \frac{3}{4}  \frac{C_{B}}{m_u^2}-  \frac{3}{4}  \frac{C_{B}}{m_u m_s}
-  \frac{3}{4}  \frac{C_{B}}{m_u m_c}
\nonumber \\ && +\frac{3}{4} \frac{C_{B}}{m_u^2} + \frac{3}{4} \frac{C_B}{m_u m_s}\nonumber \\
& =& -  \frac{3}{4}  \frac{C_{B}}{m_u m_c}
. \label{mass-diff3}
\end{eqnarray}
The same binding energy is also obtained for the other isospin partners. In another view, it can be regarded that the binding energy of these $H_c$ dibaryons is supplied from the binding energy of $(uc)$ or $(dc)$ diquarks.
As a result, one obtains $B_{H_c}=-29$ MeV and a stable $H_c$ dibaryon.  This is a new prediction and could be searched in future heavy ion collisions, where large number of charm quarks are expected to be produced and weakly decaying particle identifiable through vertex detectors \cite{Lee:2007tn}.
As further possible bound states, $(ud)(us)(sc)$ and $(ud)(ds)(sc)$ in $I=1/2$ and $(us)(ds)(sc)$ in $I=0$ all have binding energies of $-17$ MeV.
This binding energy is supplied from the binding of the $(sc)$ diquark, and hence is smaller than that of $H_{c}$ dibaryon.
Nevertheless, these configurations are additional candidates for stable dibaryons containing a single charm.

The dibaryons stated above have relatively large binding energies as the $H$ dibaryon. The diquark model also predicts other possible dibaryon states which have smaller binding energies.
In the $(qq)(qq)(qQ)$ type, we obtain stable configurations for $Q=b$ ($H_{b}$) with binding energies equal to $-9$ MeV for $I=1$, and $-5$ MeV for $I=1/2$ and for $I=0$, as shown in Table \ref{dibaryon}.
The binding energy is reduced in comparison with $Q=c$, simply because the $(ub)$, $(db)$ and $(sb)$ diquarks are more loosely bound.
As for the dibaryons containing two heavy quarks, the $(qq)(qQ)(qQ')$ and $(qq)(qq)(QQ')$ types with $Q=c$ and $Q'=b$ ($H_{bc}$ and $H'_{bc}$) can also be stable.
For the former, the binding energies are obtained as $-9$ MeV for $I=1$ and $I=3/2$, and $-5$ MeV for $I=1/2$ which bindings are supplied from each of $(ub)$, $(db)$ and $(sb)$ diquarks.
For the latter, the binding energies are $-1.8$ MeV for $I=1/2$ and $I=0$ which are supplied from the $(cb)$ diquark.
The binding energy of $(cb)$ diquark is very small in comparison with other diquarks containing light flavors.
Therefore, it supplies only small binding energy for the dibaryons.
Lastly, we can have dibaryons containing three heavy quarks as $(qq)(qQ)(QQ')$ with $Q=c$ and $Q'=b$ ($H_{bcc}$). However, their binding energies are also small because they are supplied from the $(cb)$ diquark.

According to the $SU(3)_{f}$ light flavor symmetry, the dibaryons in Table \ref{dibaryon} belong to the following multiplets.
Noting $(qq)$ diquark belongs to ${\bf \bar{3}}_{f}$, the light flavor multiplets of the $(qq)(qq)(qq)$ type are obtained by
$
{\bf \bar{3}}_{f} \times {\bf \bar{3}}_{f} \times {\bf \bar{3}}_{f} = {\bf 1}_{f} + {\bf 8}_{f} + {\bf 8}_{f} + {\bf \overline{10}}_{f}.
$
Then the $H$ dibaryon $(ud)(us)(ds)$ belongs to ${\bf 1}_{f}$.

\begin{figure}
\begin{center}
\resizebox{0.34\textwidth}{!}{%
  \includegraphics{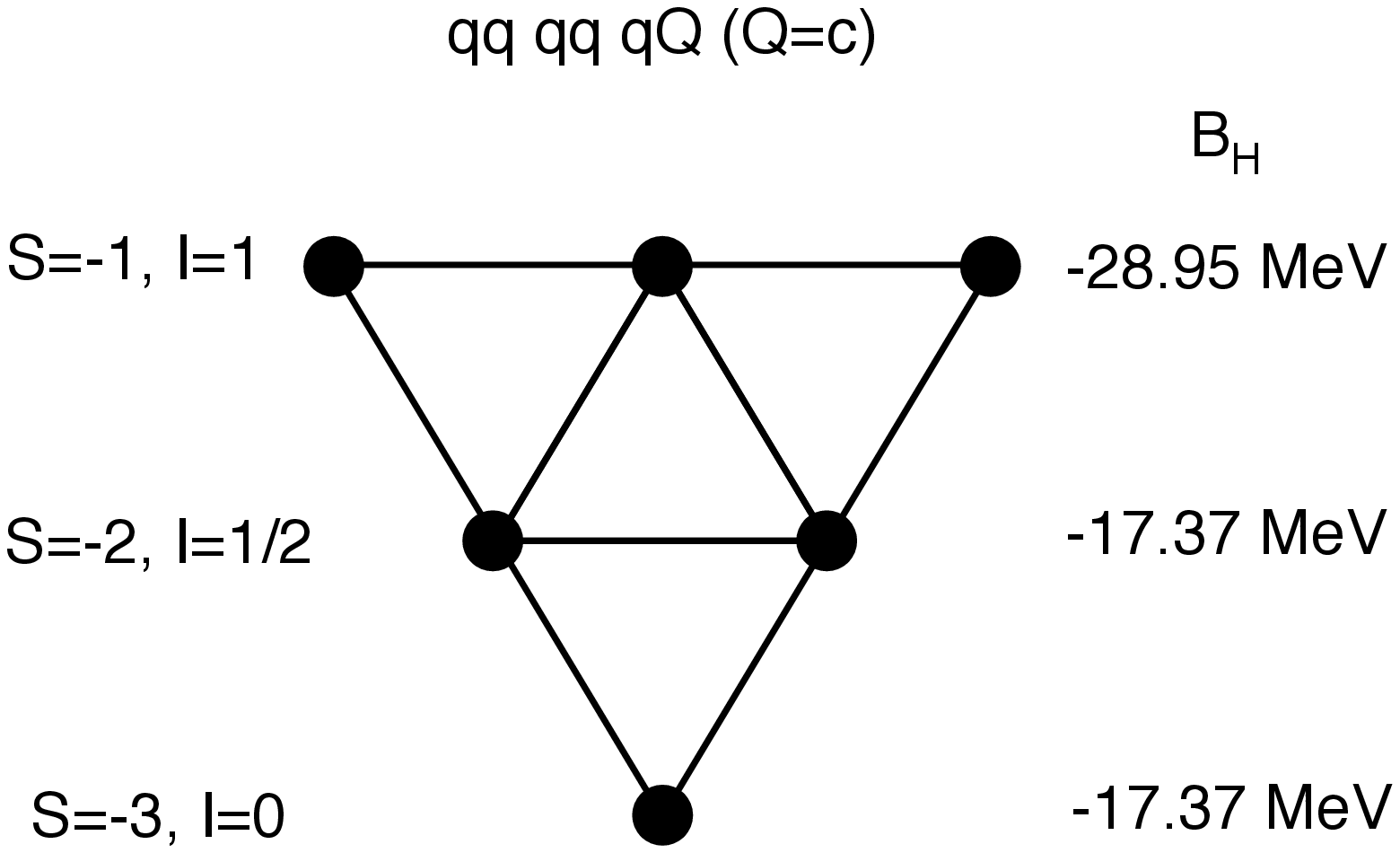}
}
\end{center}
\caption{The weight diagram of $SU(3)_{f}$ for stable diquark of $qq\,qq\,qQ$ ($Q\!=\!c$) type ($H_{c}$) with strangeness $S$, isospin $I$ and binding energy $B_{H}$. The black blobs indicate the stable dibaryons.}
\label{figure_dibaryon1}       
\end{figure}

\begin{figure}
\begin{center}
\resizebox{0.32\textwidth}{!}{%
  \includegraphics{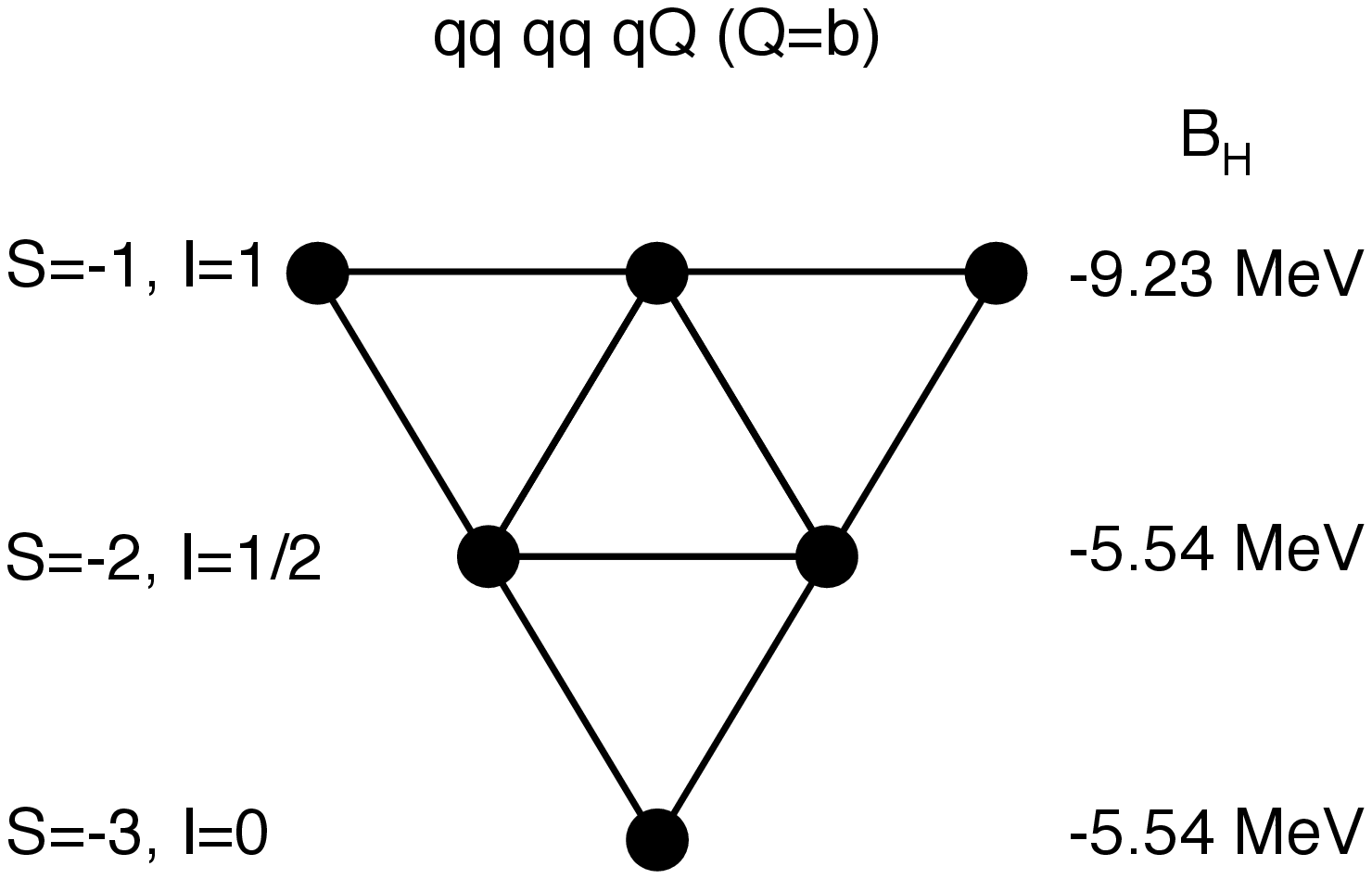}
}
\end{center}
\caption{The weight diagram of $SU(3)_{f}$ for stable diquark of $qq\,qq\,qQ$ ($Q\!=\!b$) type ($H_{b}$) with strangeness $S$, isospin $I$ and binding energy $B_{H}$. The black blobs indicate the stable dibaryons.}
\label{figure_dibaryon2}       
\end{figure}

For $(qq)(qq)(qQ)$ with $Q=c$ or $b$ ($H_{c}$ or $H_{b}$), noting $(qQ)$ diquark belongs to ${\bf 3}_{f}$, the light flavor is decomposed as
\begin{eqnarray*}
{\bf \bar{3}}_{f} \times {\bf \bar{3}}_{f} \times {\bf 3}_{f} = ({\bf 3}^{a}_{f}+{\bf \bar{6}}^{s}_{f}) \times {\bf 3}_{f}
                                                                                      = {\bf \bar{3}}^{a}_{f} + {\bf 6}_{f} + {\bf \bar{3}}^{s}_{f} + {\bf \overline{15}}_{f}.
\end{eqnarray*}
Here $a$ $(s)$ means anti-symmetry (symmetry) for a pair of two diquarks $(qq)$'s in $(qq)(qq)(qQ)$.
Because of the anti-symmetry of color in the wave function in Eq.~(\ref{H-wave}), ${\bf \bar{3}}^{s}_{f}$ and ${\bf \overline{15}}_{f}$ which contain identical pair of diquarks in symmetric flavor ${\bf \bar{6}}^{s}_{f}$ vanish, and only ${\bf \bar{3}}^{a}_{f}$ and ${\bf 6}_{f}$ which contain anti-symmetric flavor ${\bf 3}^{a}_{f}$ survive.
Concerning mixing of ${\bf \bar{3}}^{a}_{f}$ and ${\bf 6}_{f}$, some comments are in order.
First, in $S=-1$ sector, the wave function in the $I=0$ configuration in ${\bf \bar{3}}^{a}_{f}$ is given as
\begin{eqnarray*}
(ud) \left\{ (us)(dQ)-(ds)(uQ) \right\},
\end{eqnarray*}
while that of $I_{z}=0$ in $I=1$ in ${\bf 6}_{f}$, which weight vector is on the same place, is given as
\begin{eqnarray*}
(ud) \left\{ (us)(dQ)+(ds)(uQ) \right\},
\end{eqnarray*}
with appropriate normalization constants.
As far as the isospin is conserved exactly, these two states are independent.
The latter dibaryon with $I=1$ is stable against the decay to $\Lambda+\Lambda_{Q}$ because isospin symmetry forbids this decay.
However, the former dibaryon with $I=0$ is not stable, but decays to $\Lambda+\Lambda_{Q}$.
Therefore, in $S=-1$ sector, we have stable dibaryons in $I=1$.
Second, in $S=-2$ sector, the $I=1/2$ configuration in ${\bf \bar{3}}^{a}_{f}$ are generally mixed with those in ${\bf 6}_{f}$.
For example, the $I_{z}=1/2$ components are given by
\begin{eqnarray*}
(ud)(us)(sQ)\mp(us)(ds)(uQ)
\end{eqnarray*}
for ${\bf \bar{3}}^{a}_{f}$ and ${\bf 6}_{f}$, respectively.
However, from the rule to obtain the stable dibaryon, the dibaryon $(us)(ds)(uQ)$ is not stable against $\Lambda+\Xi_{Q}$ and $\Lambda_{Q}+\Xi$.
Therefore, by considering a superposition of ${\bf \bar{3}}^{a}_{f}$ and ${\bf 6}_{f}$ and eliminating $(us)(ds)(uQ)$, we obtain the stable dibaryon $(ud)(us)(sQ)$.
Third, in $S=-3$ sector, the $(us)$ and $(ds)$ pair in $(us)(ds)(sQ)$ is allowed to be $I=0$ from anti-symmetry of color configuration.
The resulting weight diagrams are shown in Fig.~\ref{figure_dibaryon1} and \ref{figure_dibaryon2} for $Q=c$ and $b$, respectively.

\begin{figure}
\begin{center}
\resizebox{0.4\textwidth}{!}{%
  \includegraphics{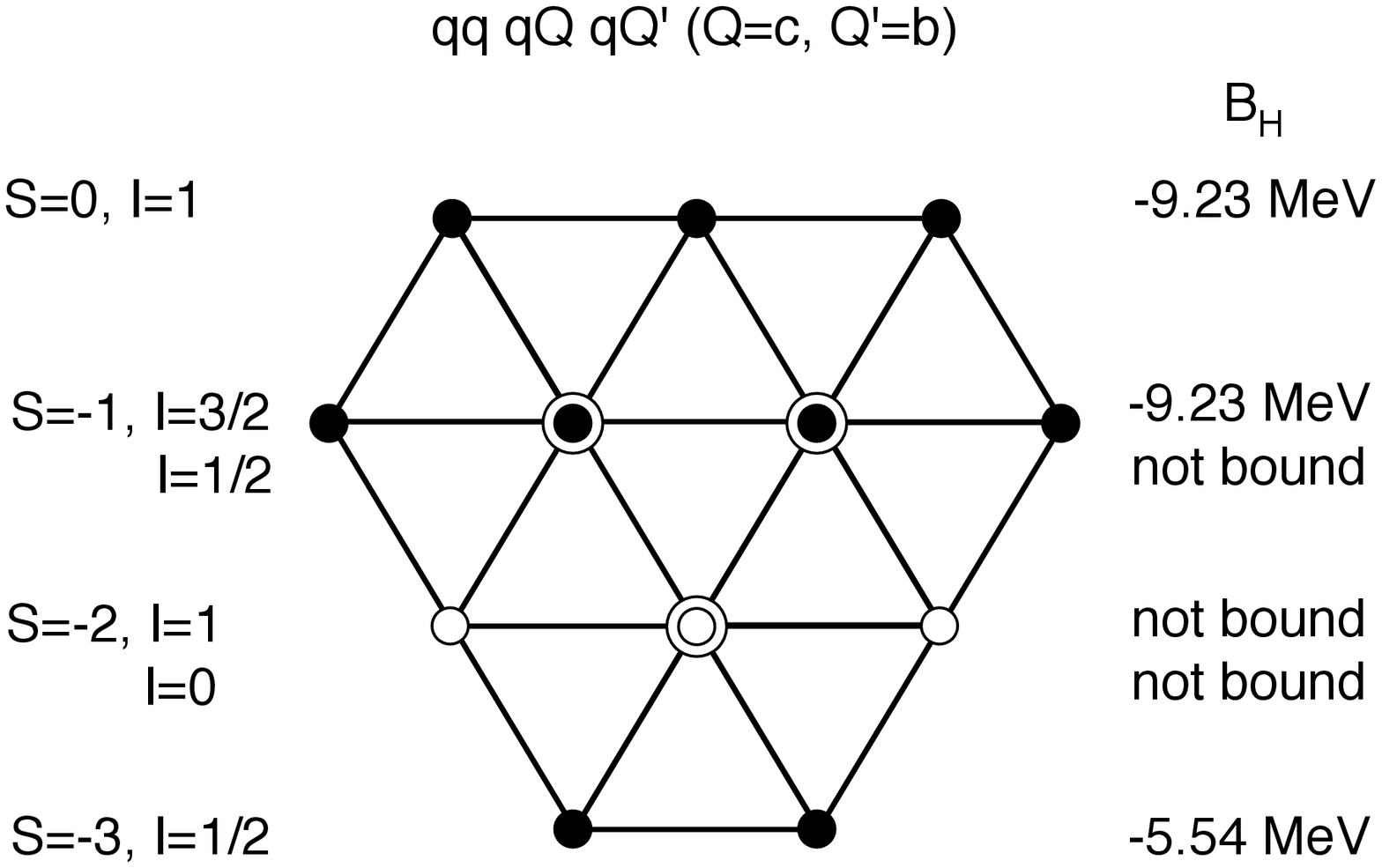}
}
\end{center}
\caption{The weight diagram of $SU(3)_{f}$ for stable diquark of $qq\,qQ\,qQ'$ ($Q\!=\!c$ and $Q'\!=\!b$) type ($H_{bc}$) with strangeness $S$, isospin $I$ and binding energy $B_{H}$. The black blobs indicate the stable dibaryons, while the white blobs do the unstable dibaryons.}
\label{figure_dibaryon3}       
\end{figure}

Similarly, other multiplets for the configurations with two or three heavy quarks are obtained.
As for $(qq)(qQ)(qQ')$ with $Q=c$ and $Q'=b$ ($H_{bc}$), the light flavor is decomposed by
\begin{eqnarray*}
{\bf \bar{3}}_{f} \times {\bf 3}_{f} \times {\bf 3}_{f} = {\bf \bar{3}}_{f} \times ( {\bf \bar{3}}^{a}_{f} + {\bf 6}^{s}_{f} ) = {\bf 3}^{a}_{f} + {\bf \bar{6}}_{f} + {\bf 3}^{s}_{f} + {\bf 15}_{f}.
\end{eqnarray*}
We note that  ${\bf 3}^{s}_{f}$ and ${\bf 15}_{f}$ containing symmetric flavor ${\bf 6}^{s}_{f}$ can survive, because $(qQ)$ cannot be identical to $(qQ')$ due to $Q \neq Q'$.
From the configurations in Table \ref{dibaryon}, we find that the stable dibaryons contain pairs of $(qQ)$ and $(qQ')$ with identical $q$, which are assigned to ${\bf 6}^{s}_{f}$.
Hence, these dibaryons belong to ${\bf 3}^{s}_{f}$ or ${\bf 15}_{f}$.
Here, we show that only ${\bf 15}_{f}$ is relevant to the stable dibaryons.
First, in $S=-1$ sector, the $I=1/2$ configuration in ${\bf 3}^{s}_{f}$ is at the same position in the weight diagram with the $I_{z} = \pm 1/2$ components in the $I=3/2$ configuration in ${\bf 15}_{f}$.
However, the $I=1/2$ configuration is not stable, because,
for example, the $I_{z}=1/2$ component in $I=1/2$ can decay to $\Lambda_{b}+\Xi_{c}^{+}$ (or $\Lambda_{c}+\Xi_{b}^{0}$).
Hence the $S=-1$ sector in ${\bf 3}^{s}_{f}$ is irrelevant.
Second, in $S=-2$ sector, any components in $I=0$ in ${\bf 3}^{s}_{f}$ and in $I=1$ in ${\bf 15}_{f}$ are not stable.
In fact we have already discussed that $(ud)(sc)(sb)$ configuration is not stable.
Therefore, ${\bf 3}^{s}_{f}$ can be discarded, and so can the $I=1$ configuration in $S=-2$ in ${\bf 15}_{f}$.
Concerning the $S=0$ sector, we note that the $I_{z}=0$ component of $I=1$  cannot decay to $\Lambda_{c}+\Lambda_{b}$ because of isospin conservation, and hence can be stable.
The resulting weight diagram is shown in Fig.~\ref{figure_dibaryon3}.
Here, the black blobs indicate the stable dibaryons, while the white ones indicate the unstable ones.

\begin{figure}
\begin{center}
\resizebox{0.28\textwidth}{!}{%
  \includegraphics{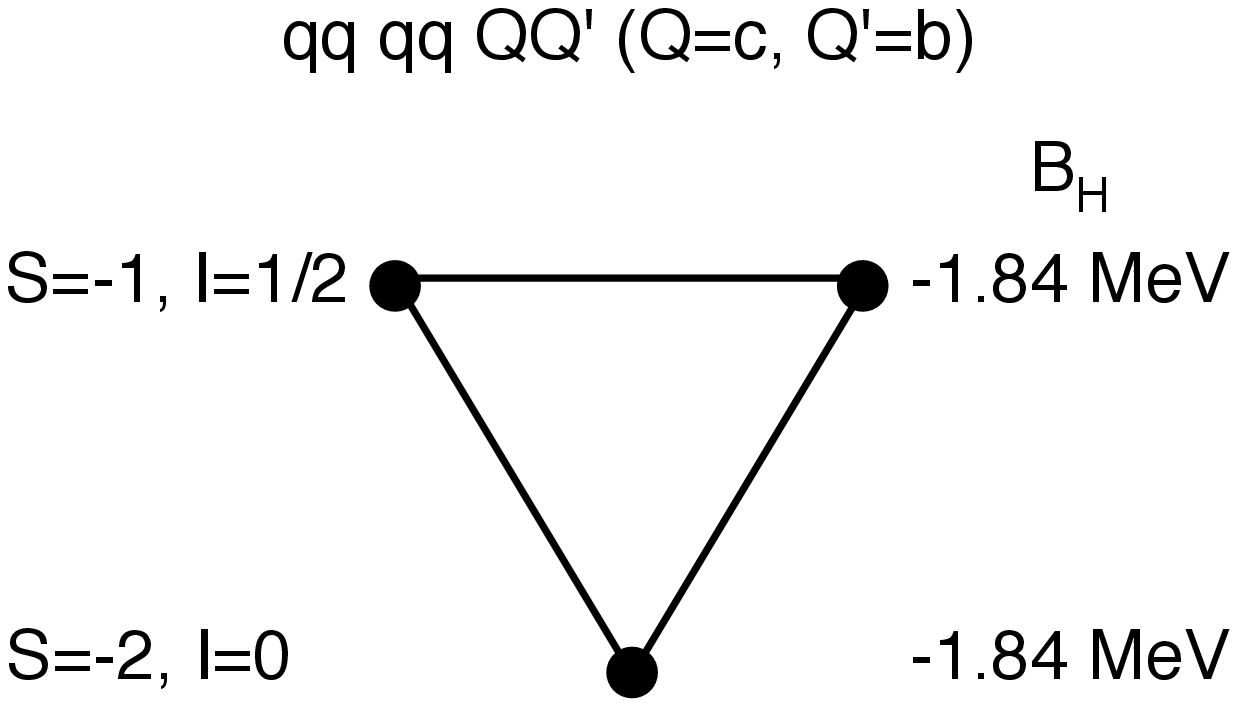}
}
\end{center}
\caption{The weight diagram of $SU(3)_{f}$ for stable diquark of $qq\,qq\,QQ'$ ($Q\!=\!c$ and $Q'\!=\!b$) type ($H'_{bc}$) with strangeness $S$, isospin $I$ and binding energy $B_{H}$. The black blobs indicate the stable dibaryons.}
\label{figure_dibaryon4}       
\end{figure}

As for the $(qq)(qq)(QQ')$ type with $Q=c$ and $Q'=b$ ($H'_{bc}$), the light flavor is decomposed as
\begin{eqnarray*}
{\bf \bar{3}}_{f} \times {\bf \bar{3}}_{f} = {\bf 3}^{a}_{f} + {\bf \bar{6}}^{s}_{f}.
\end{eqnarray*}
Because a pair of two $(qq)$'s in ${\bf \bar{6}}^{s}_{f}$ vanishes due to anti-symmetry in color, only ${\bf 3}^{a}_{f}$ can survive because of isospin conservation.
The weight diagram of  $(qq)(qq)(QQ')$ is shown in Fig.~\ref{figure_dibaryon4}.

\begin{figure}
\begin{center}
\resizebox{0.36\textwidth}{!}{%
  \includegraphics{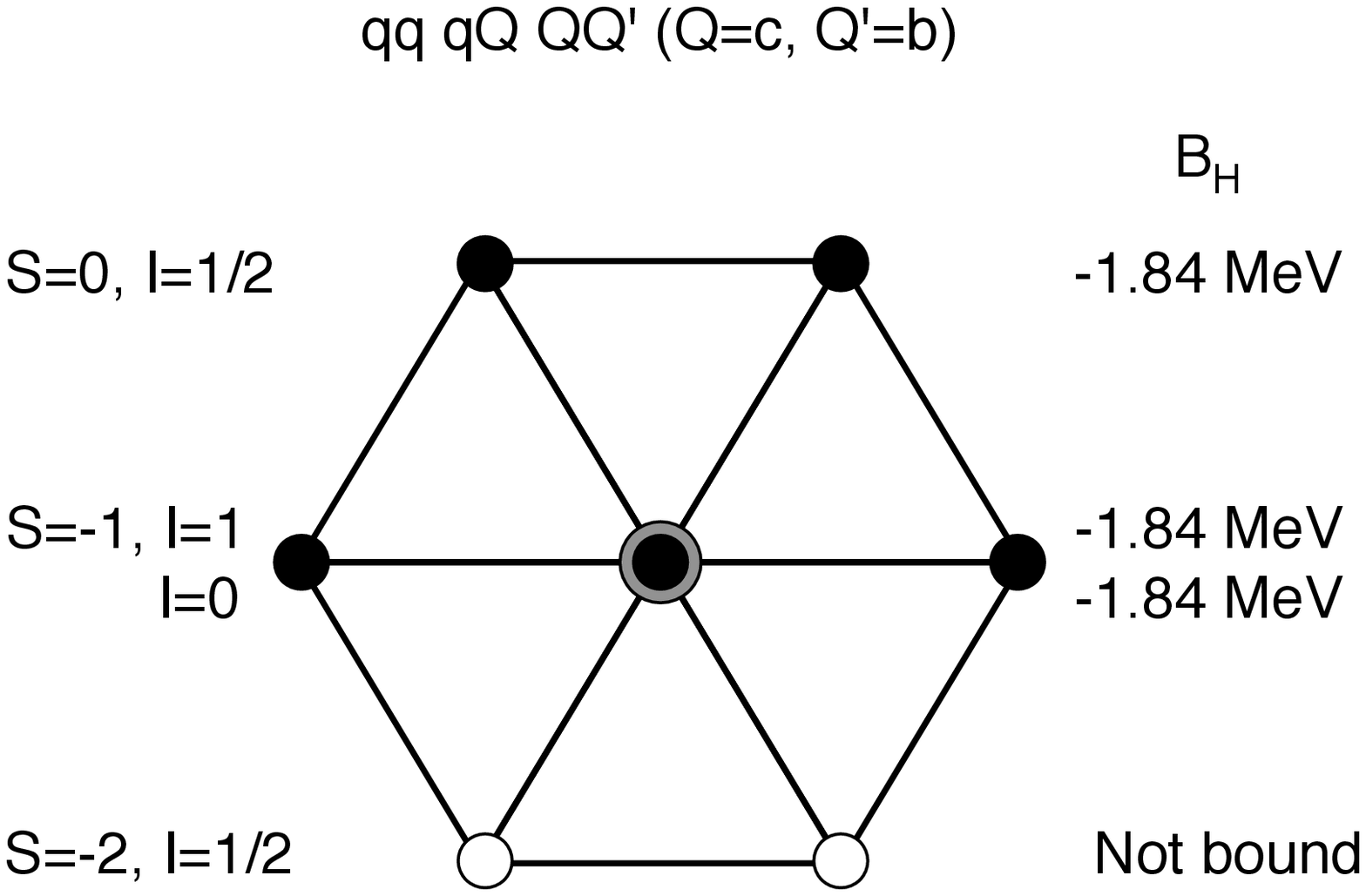}
}
\end{center}
\caption{The weight diagram of $SU(3)_{f}$ for stable diquark of $qq\,qQ\,QQ'$ ($Q\!=\!c$ and $Q'\!=\!b$) type ($H_{bcc}$) with strangeness $S$, isospin $I$ and binding energy $B_{H}$. The black blobs indicate the stable dibaryons, while the white blobs do the unstable dibaryons.}
\label{figure_dibaryon5}       
\end{figure}

Lastly, for dibaryons with three heavy quarks such as $(qq)(qQ)(QQ')$ type with $Q=c$ and $Q'=b$ ($H_{bcc}$), the light flavor is decomposed as
\begin{eqnarray*}
{\bf \bar{3}}_{f} \times {\bf 3}_{f} = {\bf 1}_{f} + {\bf 8}_{f}.
\end{eqnarray*}
In $S=-1$ sector, the $I=0$ configurations in ${\bf 1}_{f}$ are superposition of $((us)(dQ)-(ds)(uQ))(QQ')$ and $(ud)(sQ)(QQ')$, which are constructed from $(us)$, $(ds)$ and $(ud)$ diquarks.
Among them, however, the dibaryons containing $(us)$ or $(ds)$ are not stable against $\Lambda_{b}+\Omega^{+}_{cc}$ (or $\Lambda_{c}+\Omega_{cb}^{0}$), and hence $(ud)(sQ)(QQ')$ containing only $(ud)$ diquark is stable.
On the other hand, the $I=1$ configuration of $((us)(dQ)+(ds)(uQ))(QQ')$ in ${\bf 8}_{f}$ can survive.
In $S=-2$, there are no stable dibaryons.
As a result, the weight diagram is obtained in Fig.~\ref{figure_dibaryon5}.

Now, let us discuss possible decay modes from these diabryons for experiments.
Concerning $H$ dibaryon, the observation has been studied by weak decay, for example, by $H \rightarrow \Lambda + \Lambda \rightarrow p \pi^{-} + p \pi^{-}$.
Similarly, we investigate possible decay modes of the stable dibaryon including heavy quarks.

The $(qq)(qq)(qQ)$ with $Q=c$, which would be most stable state, can have the decay modes as summarized in Table \ref{decay_dibaryon1}.
For example, the wave function of the $(ud)(us)(uc)$ with $I_{z}=1$ in $I=1$ would have an overlap with $\Lambda_{c}^{+}+\Sigma^{+}$ and $p+\Xi_{c}^{+}$ with invariant mass 3475 MeV and 3406 MeV, respectively.
Then, each component will decay by weak interaction; $\Lambda_{c}^{+}+\Sigma^{+} \rightarrow pK^{-}\pi^{+}+p\pi^{0}$, and $p+\Xi_{c}^{+} \rightarrow p+\Lambda K^{-}\pi^{+}\pi^{+}$, $\Sigma^{+}K^{-}\pi^{+}$, $\Xi^{0}\pi^{+}$, $\Xi^{-}\pi^{+}\pi^{+}$, $\Xi^{0}\pi^{+}\pi^{0}$, $\Xi^{0}\pi^{+}\pi^{+}\pi^{-}$ \cite{Amsler:2008zz}.
We note that, concerning $\Lambda_{c}$, we indicate only $\Lambda_{c} \rightarrow p K^{-} \pi^{+}$ as a simplest decay pattern.
The $I_{z}=-1$ component  in $I=1$ has also an overlap with $\Lambda_{c}^{+}+\Sigma^{-}$ and $n+\Xi_{c}^{0}$ with invariant mass 3483 MeV and 3410 MeV, respectively, and the decay modes of each component are $\Lambda_{c}^{+}+\Sigma^{-} \rightarrow pK^{-}\pi^{+}+n\pi^{-}$, and $n+\Xi_{c}^{0} \rightarrow n+pK^{-}K^{-}\pi^{+}$, $\Lambda K^{-} \pi^{+}\pi^{+}\pi^{-}$, $\Xi^{-} \pi^{+}$, $\Xi^{-} \pi^{+} \pi^{+} \pi^{-}$, $\Omega^{-}K^{+}$.
The $I_{z}=0$ component in $I=1$ would contain $\Lambda+\Sigma_{c}^{+}$ or $\Lambda_{c}^{+}+\Sigma^{-}$ because of isospin symmetry, and hence its decay modes are $\Lambda+\Sigma_{c}^{+} \rightarrow p\pi^{-}+\Lambda_{c}\pi^{0} \rightarrow p\pi^{-}+pK^{-}\pi^{+}\pi^{0}$, and $\Lambda_{c}^{+}+\Sigma^{-} \rightarrow \Lambda \gamma + pK^{-}\pi^{+}$.
However, in reality, explicit isospin breaking effect would induce strong decay much larger than weak decay.
Hence these states would first decay to $H_c \rightarrow \Lambda+\Lambda_{c}^{+}$ by explicit isospin breaking, and second they would decay to $p \pi^{-}+pK^{-}\pi^{+}\pi^{0}$ by weak interaction.
As for $I=1/2$ and $I=0$, which would be second most stable states, we obtain the possible decay modes as shown in \ref{decay_dibaryon1}.
Note that the isospin $I=0$ indicates the wave function $\Xi^{0}\Xi_{c}^{0} - \Xi^{-}\Xi_{c}^{+}$ and each component is shown in the table.

Similar decay patters hold for $(qq)(qq)(qQ)$ with $Q=b$ as shown in Table \ref{decay_dibaryon2}.
However, because of the lack of experimental data about the decays of several bottom baryons, available decay patterns are limited at present.
Nevertheless at least the $I=1$ and $I=1/2$ dibaryons can be detectable, while the $I=0$ dibaryon needs experimental information about the decays of $\Xi_{b}^{0}$ or $\Xi_{b}^{-}$.

As for dibaryons with two or three heavy quarks, only a few candidates are available.
In $(qq)(qQ)(qQ')$ with $Q=c$ and $Q'=b$, the $I_{z}=\pm1$ components in $I=1$ and $I=3/2$ could be investigated by using experimentally known decay patterns as shown in Table \ref{decay_dibaryon3}.
In $(qq)(qq)(QQ')$ with $Q=c$ and $Q'=b$, the $I_{z}=\pm1/2$ components  in $I=1/2$ could be also investigated as shown in Table \ref{decay_dibaryon4}.
Lastly, any components in $(qq)(qQ)(QQ')$ with $Q=c$ and $Q'=b$ would be difficult because of the lack of experimental information as shown in Table \ref{decay_dibaryon5}.

Thus, the diquark model gives a list of possible $H$ dibaryons, $H_{c}$, $H_{b}$, $H_{bc}$, $H_{bc}^{\prime}$ and $H_{bcc}$.
Among them, the first promising candidate is $H_{c}$ with $I=0$, and the second ones are $H_{c}$ with $I=1/2$ and 1.
The other $H$ dibaryons, $H_{b}$, $H_{bc}$, $H_{bc}^{\prime}$ and $H_{bcc}$, are less probable because of their small binding energies.
Nevertheless, search of these particles is a challenging subject, not only for theory with more sophisticated analysis  by lattice simulation, QCD sum rule, and so on, but also for experiments in high energy facilities.

\begin{table*}[p]
\caption{The mass difference of $H$-dibaryons and two baryon state $B$ and $B'$. $B_{H} =m_{H}-m_{B}-m_{B'}$. The $H$-dibaryons can be stable for $B_{H} < 0$. $Q \neq Q'$. The unit is in MeV. Contrary to the text, we omit brackets denoting scalar and isoscalar diquark for simplicity.}
\label{dibaryon}
\begin{tabular}{|c|c|c|c|c|c|c|r|}
\hline
& flavor  & config. of $H$ & $I$ & $I_{z}$ & $S$ & config. of $B+B'$ & $B_{H}$ [MeV] \\
\hline
\hline
$H$$(qq\,qq\,qq)$ & 22200 & $ud\,us\,ds$ & 0 & 0 & -2 & $ud\,s+ud\,s$ & \,-28.95 \\
\hline
\hline
           	      & 31110 & $ud\,us\,uc$ & 1 & 1 & -1 & $ud\,c+us\,u$, $ud\,u+us\,c$  &  \\
\cline{2-7}
                      & 22110 & $ud\,\frac{1}{\sqrt{2}}(us\,dc + ds\,uc)$ & 1 & 0 & -1 & $ud\,s+ud\,c$, $ud\,s+ud\,c$  & \,-28.95 \\
\cline{2-7}
$H_{c}(qq\,qq\,qc)$ & 13110 & $ud\,ds\,dc$ & 1 & -1 & -1 & $ud\,c+ds\,d$, $ud\,d+ds\,c$ &  \\
\cline{2-8}
                      & 21210 & $ud\,us\,sc$ & 1/2 & 1/2 & -2 & $ud\,c+us\,s$, $ud\,s+us\,c$ & \,-17.37 \\
\cline{2-7}
                      & 12210 & $ud\,ds\,sc$ & 1/2 & -1/2 & -2 & $ud\,c+ds\,s$, $ud\,s+ds\,c$ &  \\
\cline{2-8}
                      & 11310 & $\frac{1}{\sqrt{2}} (us\,ds-ds\,us)\,sc$ & 0 & 0 & -3 & $us\,s+ds\,c$, $us\,c+ds\,s$ & \,-17.37  \\
\hline
\hline
                      & 31101 & $ud\,us\,ub$ & 1 & 1 & -1 & $ud\,u+us\,b$, $ud\,b+us\,u$ &  \\
\cline{2-7}
                      & 22101 & $ud\,\frac{1}{\sqrt{2}}(us\,db + ds\,ub)$ & 1 & 0 & -1 & $ud\,s+ud\,b$, $ud\,s+ud\,b$ & \,\,-9.23 \\
\cline{2-7}
$H_{b}(qq\,qq\,qb)$ & 13101 & $ud\,ds\,db$ & 1 & -1 & -1 & $ud\,b+ds\,d$, $ud\,d+ds\,b$ &  \\
\cline{2-8}
                      & 21202 & $ud\,us\,sb$ & 1/2 & 1/2 & -2 & $ud\,b+us\,s$, $ud\,s+us\,b$ & \,\,-5.54 \\
\cline{2-7}
                      & 12201 & $ud\,ds\,sb$ & 1/2 & -1/2 & -2 & $ud\,b+ds\,s$, $ud\,s+ds\,b$ & \\
\cline{2-8}
                      & 11301 & $\frac{1}{\sqrt{2}} (us\,ds-ds\,us)\,sb$ & 0 & 0 & -3 & $us\,b+ds\,s$, $us\,s+ds\,b$ & \,\,-5.54 \\
\hline
\hline
                       & 31011 & $ud\,uc\,ub$ & 1 & 1 & 0 & $ud\,b+uc\,u$, $ud\,u+uc\,b$ & \\
\cline{2-7}
                       & 22011 & $ud\,\frac{1}{\sqrt{2}}(uc\,db + dc\,ub)$ & 1 & 0 & 0 &$ud\,b+uc\,u$, $ud\,u+uc\,b$ & \,\,-9.23 \\
\cline{2-7}
                       & 13011 & $ud\,dc\,db$ & 1 & -1 & 0 & $ud\,b+dc\,d$, $ud\,d+dc\,b$ & \\
\cline{2-8}
                       & 30111 & $us\,uc\,ub$ & 3/2 & 3/2 & -1 & $us\,u+uc\,b$, $us\,b+uc\,u$ &  \\
\cline{2-7}
$H_{bc}(qq\,qc\,qb)$ & 21111 & $\frac{1}{\sqrt{3}}( (us\,dc + ds\,uc)\,ub + us\,uc\,db)$ & 3/2 & 1/2 & -1 & $ud\,b+us\,c$, $ud\,c+us\,b$ & \,\,-9.23 \\
\cline{2-7}
                       & 12111 & $\frac{1}{\sqrt{3}}( (us\,dc + ds\,uc)\,db + ds\,dc\,ub)$ & 3/2 & -1/2 & -1 &$ud\,b+ds\,c$, $ud\,c+ds\,b$ &  \\
\cline{2-7}
                       & 03111 & $ds\,dc\,db$ & 3/2& -3/2 & -1 & $ds\,b+dc\,d$, $ds\,d+dc\,b$ &  \\
\cline{2-8}
                       & 10311 & $us\,sc\,sb$ & 1/2 & 1/2 & -3 &$us\,b+sc\,s$, $us\,s+sc\,b$ & \,\,-5.54 \\
\cline{2-7}
                       & 01311 & $ds\,sc\,sb$ & 1/2 & -1/2 & -3 & $ds\,b+sc\,s$, $ds\,s+sc\,b$ &  \\
\hline
\hline
                       & 21111 & $ud\,us\,cb$ & 1/2 & 1/2 & -1 & $ud\,b+us\,c$, $ud\,c+us\,b$  & \,\,-1.84 \\
\cline{2-7}
$H'_{bc}(qq\,qq\,bc)$ & 12111 & $ud\,ds\,cb$ & 1/2 & -1/2 & -1 & $ud\,b+ds\,c$, $ud\,c+ds\,b$  &  \\
\cline{2-8}
                       & 11211 & $\frac{1}{\sqrt{2}} (us\,ds-ds\,us)\,cb$ & 0 & 0 & -2 & $us\,b+ds\,c$, $us\,c+ds\,b$  & \,-1.84 \\
\hline
\hline
                        & 21021 & $ud\,uc\,cb$ & 1/2 & 1/2 & 0 & $ud\,b+uc\,c$, $ud\,c+uc\,b$  & \,-1.84 \\
\cline{2-7}
$$           & 12021 & $ud\,dc\,cb$ & 1/2 & -1/2 & 0 & $ud\,b+dc\,c$, $ud\,c+dc\,b$ & \\
\cline{2-8}
$H_{bcc}(qq\,qc\,cb)$ & 20121 & $us\,uc\,cb$ & 1 & 1 & -1 & $us\,b+uc\,c$, $us\,c+uc\,b$ & \\
\cline{2-7}
                        & 11121 & $\frac{1}{\sqrt{2}}(us\,dc + ds\,uc)\,cb$ & 1 & 0 & -1 & $us\,b+dc\,c$, $us\,c+dc\,b$ & \,-1.84 \\
\cline{2-7}
                        & 02121 & $ds\,dc\,cb$ & 1 & -1 & -1 & $ds\,b+dc\,c$, $ds\,c+dc\,b$ & \\
\cline{2-8}
                        & 11121 & $ud\,sc\,cb$ & 0 & 0 & -1 & $ud\,b+sc\,c$, $ud\,c+sc\,b$  & \,-1.84 \\
\hline
\end{tabular}
\end{table*}%

\begin{table*}[p]
\caption{[$qq\,qq\,qQ$; $Q\!=\!c$ ($H_{c}$)] The possible decay modes of $H_{c}$-dibaryons with configuration $qq\,qq\,qQ$ ($Q\!=\!c$) against two baryon state $B$ and $B'$. The numbers in the brackets are invariant mass of the two baryons at rest frame.}
\begin{center}
\begin{tabular}{|c|c|c|c|c|c|c|}
\hline
flavor & config. of $H_{c}$ & $I$ & $I_{z}$ & $S$ & config. of $B+B'$ (MeV) & final states \\
\hline
31110 & $ud\,us\,uc$ & 1 & 1 & -1 & $\Lambda_{c}^{+}+\Sigma^{+}$ (3475.83)  & $pK^{-}\pi^{+}+p \pi^{0}$  \\
\cline{6-7}
 & & & & & $p+\Xi_{c}^{+}$ (3406.17) & $p+\Lambda K^{-}\pi^{+}\pi^{+}$ \\
 & & & & &                                                       & $p + \Sigma^{+}K^{-}\pi^{+}$ \\
 & & & & &                                                       & $p + \Xi^{0}\pi^{+}$ \\
 & & & & &                                                       & $p + \Xi^{-}\pi^{+}\pi^{+}$ \\
 & & & & &                                                       & $p + \Xi^{0}\pi^{+}\pi^{0}$ \\
 & & & & &                                                       & $p + \Xi^{0}\pi^{+}\pi^{+}\pi^{-}$ \\
\hline
22110 & $ud\,\frac{1}{\sqrt{2}}(us\,dc + ds\,uc)$ & 1 & 0 & -1 & $\Lambda+\Sigma_{c}^{+}$ (3568.58)  & $p\pi^{-}+\Lambda_{c}^{+}\pi^{0}$ \\
\cline{6-7}
 & & & & & $\Sigma^{0}+\Lambda_{c}^{+}$ (3478.92) & $\Lambda \gamma + p K^{-} \pi^{+}$ \\
\hline
13110 & $ud\,ds\,dc$ & 1 & -1 & -1 & $\Lambda_{c}^{+}+\Sigma^{-}$ (3483.91) & $pK^{-}\pi^{+}+n\pi^{-}$ \\
\cline{6-7}
 & & & & & $n+\Xi_{c}^{0}$ (3410.56) & $n+pK^{-}K^{-}\pi^{+}$ \\
 & & & & &                                           & $n+\Lambda K^{-} \pi^{+}\pi^{+}\pi^{-}$ \\
 & & & & &                                          & $n+\Xi^{-} \pi^{+}$ \\
 & & & & &                                          & $n+\Xi^{-} \pi^{+} \pi^{+} \pi^{-}$ \\
 & & & & &                                          & $n+\Omega^{-}K^{+}$ \\
\hline
21210 & $ud\,us\,sc$ & 1/2 & 1/2 & -2 & $\Lambda_{c}^{+}+\Xi^{0}$ (3601.32) & $pK^{-}\pi^{+}+\Lambda \pi^{0}$ \\
\cline{6-7}
 & & & & & $\Lambda+\Xi_{c}^{+}$ (3583.58) & $p \pi^{-}+\Lambda K^{-}\pi^{+}\pi^{+}$ \\
 & & & & &                                                       & $p \pi^{-}+\Sigma^{+}K^{-}\pi^{+}$ \\
 & & & & &                                                       & $p \pi^{-}+\Xi^{0}\pi^{+}$ \\
 & & & & &                                                       & $p \pi^{-}+\Xi^{-}\pi^{+}\pi^{+}$ \\
 & & & & &                                                       & $p \pi^{-}+\Xi^{0}\pi^{+}\pi^{0}$ \\
 & & & & &                                                       & $p \pi^{-}+\Xi^{0}\pi^{+}\pi^{+}\pi^{-}$ \\
\hline
12210 & $ud\,ds\,sc$ & 1/2 & -1/2 & -2 & $\Lambda_{c}^{+}+\Xi^{-}$ (3608.17) & $pK^{-}\pi^{+}+\Lambda \pi^{-}$ \\
\cline{6-7}
 & & & & & $\Lambda+\Xi_{c}^{0}$ (3586.68) & $p \pi^{-}+pK^{-}K^{-}\pi^{+}$ \\
 & & & & &                                                       & $p \pi^{-}+\Lambda K^{-}\pi^{+}\pi^{+}\pi^{-}$ \\
 & & & & &                                                       & $p \pi^{-}+\Xi^{-}\pi^{+}$ \\
 & & & & &                                                       & $p \pi^{-}+\Xi^{-}\pi^{+}\pi^{+}\pi^{-}$ \\
 & & & & &                                                       & $p \pi^{-}+\Omega^{-}K^{+}$ \\
\hline
11310 & $us\,ds\,sc$ & 0 & 0 & -3 & $\Xi^{0}+\Xi_{c}^{0}$ (3785.86) & $\Lambda \pi^{0} + pK^{-}K^{-}\pi^{+}$ \\
 & & & & &                                                      & $\Lambda \pi^{0} + \Lambda K^{-}\pi^{+}\pi^{+}\pi^{-}$ \\
 & & & & &                                                      & $\Lambda \pi^{0} + \Xi^{-}\pi^{+}$ \\
 & & & & &                                                      & $\Lambda \pi^{0} + \Xi^{-}\pi^{+}\pi^{+}\pi^{-}$ \\
 & & & & &                                                      & $\Lambda \pi^{0} + \Omega^{-}K^{+}$ \\
\cline{6-7}
 & & & & & $\Xi^{-}+\Xi_{c}^{+}$ (3789.6) & $\Lambda \pi^{-} + \Lambda K^{-} \pi^{+} \pi^{+}$ \\
 & & & & &                                                 & $\Lambda \pi^{-} + \Sigma^{+}K^{-}\pi^{+}$ \\
 & & & & &                                                 & $\Lambda \pi^{-} + \Xi^{0}\pi^{+}$ \\
 & & & & &                                                 & $\Lambda \pi^{-} + \Xi^{-}\pi^{+}\pi^{+}$ \\
 & & & & &                                                 & $\Lambda \pi^{-} + \Xi^{0}\pi^{+}\pi^{0}$ \\
 & & & & &                                                 & $\Lambda \pi^{-} + \Xi^{0}\pi^{+}\pi^{+}\pi^{-}$ \\
\hline
\end{tabular}
\end{center}
\label{decay_dibaryon1}
\end{table*}%

\begin{table*}[p]
\caption{[$qq\,qq\,qQ$; $Q\!=\!b$ ($H_{b}$)] The possible decay modes of $H_{b}$-dibaryons with configuration $qq\,qq\,qQ$ ($Q\!=\!b$) against two baryon state $B$ and $B'$. The numbers in the brackets are invariant mass of the two baryons at rest frame.}
\begin{center}
\begin{tabular}{|c|c|c|c|c|c|c|}
\hline
flavor & config. of $H_{b}$ & $I$ & $I_{z}$ & $S$ & config. of $B+B'$ (MeV) & final states \\
\hline
31101 & $ud\,us\,ub$ & 1 & 1 & -1 & $p+\Xi_{b}^{0}$ (6730.67) & $p+?$ \\
\cline{6-7}
 & & & & & $\Lambda_{b}^{0}+\Sigma^{+}$ (6809.57) & $\Lambda_{c}^{+}\pi^{-}+p\pi^{0}$ \\
\hline
22101 & $ud\,\frac{1}{\sqrt{2}}(us\,db + ds\,ub)$ & 1 & 0 & -1 & $\Lambda+\Sigma_{b}^{0}$ (6926.18) & $p\pi^{-}+\Lambda_{b}^{0}\pi^{0}$ \\
\cline{6-7}
 & & & & & $\Sigma^{0}+\Lambda_{b}^{0}$ (6812.84) & $\Lambda \gamma + \Lambda_{c}^{+}\pi^{-}$ \\
\hline
13101 & $ud\,ds\,db$ & 1 & -1 & -1 & $\Lambda_{b}^{0}+\Sigma^{-}$ (6817.0) & $\Lambda_{c}^{+}\pi^{-} + n\pi^{-}$ \\
\cline{6-7}
 & & & & & $n+\Xi_{b}^{-}$ (6731.97) & $n + ?$ \\
\hline
21202 & $ud\,us\,sb$ & 1/2 & 1/2 & -2 & $\Lambda_{b}^{0}+\Xi^{0}$ (6935.06) & $\Lambda_{c}^{+} \pi^{-} + \Lambda \pi^{0}$ \\
\cline{6-7}
 & & & & & $\Lambda + \Xi_{b}^{0}$ (6908.08) & $p \pi^{-} + ?$ \\
\hline
12201 & $ud\,ds\,sb$ & 1/2 & -1/2 & -2 & $\Lambda_{b}^{0}+\Xi^{-}$ (6941.9) & $\Lambda_{c}^{+}\pi^{-} + \Lambda \pi^{-}$ \\
\cline{6-7}
 & & & & & $\Lambda + \Xi_{b}^{-}$ (6908.08) & $p\pi^{-} + ?$ \\
\hline
11301 & $us\,ds\,sb$ & 0 & 0 & -3 & $\Xi_{b}^{0} + \Xi^{-}$ (7114.1) & $?+\Lambda \pi^{-}$ \\
\cline{6-7}
 & & & & & $\Xi^{0} + \Xi_{b}^{-}$ (7107.26) & $\Lambda \pi^{0} + ?$ \\
\hline
\end{tabular}
\end{center}
\label{decay_dibaryon2}
\end{table*}%

\begin{table*}[p]
\caption{[$qq\,qQ\,qQ'$; $Q\!=\!c$, $Q'\!=\!b$ ($H_{bc}$)] The possible decay modes of $H_{bc}$-dibaryons with configuration $qq\,qQ\,qQ'$ ($Q\!=\!c$, $Q'\!=\!b$) against two baryon state $B$ and $B'$. The numbers in the brackets are invariant mass of the two baryons at rest frame.}
\begin{center}
\begin{tabular}{|c|c|c|c|c|c|c|}
\hline
flavor & config. of $H_{bc}$ & $I$ & $I_{z}$ & $S$ & config. of $B+B'$ (MeV) & final states \\
\hline
31011 & $ud\,uc\,ub$ & 1 & 1 & 0 & $\Lambda_{b}^{0}+\Sigma_{c}^{++}$  (8074.2) & $\Lambda_{c}^{+}\pi^{-}+\Lambda_{c}^{+}\pi^{+}$ \\
\cline{6-7}
 & & & & & $p+\Xi_{cb}^{+}$ (?) & $p + ?$ \\
\hline
22011 & $ud\,\frac{1}{\sqrt{2}}(uc\,db + dc\,ub)$ & 1 & 0 & 0 & $\Xi_{cb}^{0}+p$ (?) & $?+p$ \\
\cline{6-7}
 & & & & & $n+\Xi_{cb}^{+}$ (?) & $n + ?$ \\
\hline
13011 & $ud\,dc\,db$ & 1 & -1 & 0 & $\Lambda_{b}^{0}+\Sigma_{c}^{0}$ (8073.96) & $\Lambda_{c}^{+}\pi^{-}+\Lambda_{c}^{+}\pi^{-}$ \\
\cline{6-7}
 & & & & & $n+\Xi_{cb}^{0}$ (?) & $n + ?$ \\
\hline
30111 & $us\,uc\,ub$ &3/2 & 3/2 & -1 & $\Sigma^{+}+\Xi_{cb}^{+}$ (?) & $p\pi^{0}+?$ \\
\cline{6-7}
 & & & & & $\Xi_{b}^{0}+\Sigma_{c}^{++}$ (8246.4) & $? + \Lambda_{c}^{+}\pi^{+}$ \\
\hline
21111 & $\frac{1}{\sqrt{3}}(us\,uc\,db + us\,dc\,ub + ds\,uc\,ub)$ &3/2 & 1/2 & -1 & $\Sigma_{b}^{0}+\Xi_{c}^{+}$ (8278.4) & $\Lambda_{b}^{0}\pi^{0}+\Lambda K^{-} \pi^{+} \pi^{+}$ \\
 & & & & & & $\Lambda_{b}^{0}\pi^{0}+\Sigma^{+} K^{-} \pi^{+}$ \\
 & & & & & & $\Lambda_{b}^{0}\pi^{0}+\Xi^{0} \pi^{+}$ \\
 & & & & & & $\Lambda_{b}^{0}\pi^{0}+\Xi^{-} \pi^{+} \pi^{+}$ \\
 & & & & & & $\Lambda_{b}^{0}\pi^{0}+\Xi^{0} \pi^{+} \pi^{0}$ \\
 & & & & & & $\Lambda_{b}^{0}\pi^{0}+\Xi^{0} \pi^{+} \pi^{+} \pi^{-}$ \\
\cline{6-7}
 & & & & & $\Sigma_{c}^{+}+\Xi_{b}^{0}$ (8245.3) & $\Lambda_{c}^{+}\pi^{0} + ?$ \\
\hline
12111 & $\frac{1}{\sqrt{3}}(us\,dc\,db + ds\,dc\,ub + ds\,uc\,db)$ &3/2 & -1/2 & -1 & $\Sigma_{b}^{0}+\Xi_{c}^{0}$ (8282.5) & $\Lambda_{b}^{0}\pi^{0}+pK^{-}K^{-}\pi^{+}$ \\
 & & & & & & $\Lambda_{b}^{0}\pi^{0} + \Lambda K^{-} \pi^{+} \pi^{+} \pi^{-}$ \\
 & & & & & & $\Lambda_{b}^{0}\pi^{0} + \Xi^{-} \pi^{+}$ \\
 & & & & & & $\Lambda_{b}^{0}\pi^{0} + \Xi^{-} \pi^{+} \pi^{+} \pi^{+} \pi^{-}$ \\
 & & & & & & $\Lambda_{b}^{0}\pi^{0} + \Omega^{-}K^{+}$ \\
\cline{6-7}
 & & & & & $\Sigma_{c}^{+}+\Xi_{b}^{-}$ (8309.9) & $\Lambda_{c}^{+}\pi^{0} + ?$ \\
\hline
03111 & $ds\,dc\,db$ & 3/2& -3/2 & -1 & $\Xi_{b}^{-}+\Sigma_{c}^{0}$ (8246.16) & $?+\Lambda_{c}^{+}\pi^{-}$ \\
\cline{6-7}
 & & & & & $\Sigma^{-}+\Xi_{cb}^{0}$ (?) & $n\pi^{-} + ?$ \\
\hline
10311 & $us\,sc\,sb$ & 1/2 & 1/2 & -3 & $\Xi_{b}^{0}+\Omega_{c}^{0}$ (8489.9) & $?+\Sigma^{+}K^{-}K^{-}\pi^{+}$ \\
 & & & & & & $?+\Xi^{0} K^{-} \pi^{+}$ \\
 & & & & & & $?+\Xi^{-} K^{-} \pi^{+} \pi^{+}$ \\
 & & & & & & $?+\Omega^{-} \pi^{+}$ \\
 & & & & & & $?+\Omega^{-} \pi^{+} \pi^{0}$ \\
 & & & & & & $?+\Omega^{-} \pi^{+} \pi^{+} \pi^{-}$ \\
\cline{6-7}
 & & & & & $\Xi^{0}+\Omega_{cb}^{0}$ (?) & $\Lambda\pi^{0} + ?$ \\
\hline
01311 & $ds\,sc\,sb$ & 1/2 & -1/2 & -3 & $\Xi_{b}^{-}+\Omega_{c}^{0}$ (8489.9) & $?+\Sigma^{+}K^{-}K^{-}\pi^{+}$ \\
 & & & & & & $?+\Xi^{0}K^{-}\pi^{+}$ \\
 & & & & & & $?+\Xi^{-}K^{-}\pi^{+}\pi^{+}$ \\
 & & & & & & $?+\Omega^{-}\pi^{+}$ \\
 & & & & & & $?+\Omega^{-}\pi^{+}\pi^{0}$ \\
 & & & & & & $?+\Omega^{-}\pi^{+}\pi^{+}\pi^{-}$ \\
\cline{6-7}
 & & & & & $\Xi^{-}+\Omega_{cb}^{0}$ (?) & $\Lambda\pi^{-} + ?$ \\
\hline
\end{tabular}
\end{center}
\label{decay_dibaryon3}
\end{table*}%

\begin{table*}[p]
\caption{[$qq\,qq\,QQ'$; $Q\!=\!c$, $Q'\!=\!b$ ($H'_{bc}$)] The possible decay modes of $H'_{bc}$-dibaryons with configuration $qq\,qq\,QQ'$ ($Q\!=\!c$, $Q'\!=\!b$) against two baryon state $B$ and $B'$. The numbers in the brackets are invariant mass of the two baryons at rest frame.}
\begin{center}
\begin{tabular}{|c|c|c|c|c|c|c|}
\hline
flavor & config. of $H'_{bc}$ & $I$ & $I_{z}$ & $S$ & config. of $B+B'$ (MeV) & final states \\
\hline
21111 & $ud\,us\,cb$ & 1/2 & 1/2 & -1 & $\Lambda_{b}^{0}+\Xi_{c}^{+}$ (8088.1) & $\Lambda_{c}^{+}\pi^{-}+\Lambda K^{-}\pi^{+}\pi^{+}$ \\
 & & & & & & $\Lambda_{c}^{+}\pi^{-} + \Sigma^{+} K^{-} \pi^{+}$ \\
 & & & & & & $\Lambda_{c}^{+}\pi^{-} + \Xi^{0} \pi^{+}$ \\
 & & & & & & $\Lambda_{c}^{+}\pi^{-} + \Xi^{-} \pi^{+} \pi^{+}$ \\
 & & & & & & $\Lambda_{c}^{+}\pi^{-} + \Xi^{0} \pi^{+} \pi^{0}$ \\
 & & & & & & $\Lambda_{c}^{+}\pi^{-} + \Xi^{0} \pi^{+} \pi^{+} \pi^{-}$ \\
\cline{6-7}
 & & & & & $\Lambda_{c}^{+}+\Xi_{b}^{0}$ (8078.86) & $pK^{-}\pi^{+}+?$ \\
\hline
12111 & $ud\,ds\,cb$ & 1/2 & -1/2 & -1 & $\Lambda_{b}^{0}+\Xi_{c}^{0}$ (8091.2) & $\Lambda_{c}^{+}\pi^{-}+pK^{-}K^{-}\pi^{+}$ \\
 & & & & & & $\Lambda_{c}^{+}\pi^{-} + \Lambda K^{-}\pi^{+}\pi^{+}\pi^{-}$ \\
 & & & & & & $\Lambda_{c}^{+}\pi^{-} + \Xi^{-} \pi^{+}$ \\
 & & & & & & $\Lambda_{c}^{+}\pi^{-} + \Xi^{-} \pi^{+} \pi^{+} \pi^{-}$ \\
 & & & & & & $\Lambda_{c}^{+}\pi^{-} + \Omega^{-} K^{+}$ \\
\cline{6-7}
 & & & & & $\Lambda_{c}^{+}+\Xi_{b}^{-}$ (8078.86) & $pK^{-}\pi^{+}+?$ \\
\hline
11211 & $us\,ds\,cb$ & 0 & 0 & -2 & $\Xi_{b}^{+}+\Xi_{c}^{0}$ (8263.4) & $?+pK^{-}K^{-}\pi^{+}$ \\
 & & & & & & $? + \Lambda K^{-}\pi^{+}\pi^{+}\pi^{-}$ \\
 & & & & & & $? + \Xi^{-}\pi^{+}$ \\
 & & & & & & $? + \Xi^{-}\pi^{+}\pi^{+}\pi^{-}$ \\
 & & & & & & $? + \Omega^{-} K^{+}$ \\
\cline{6-7}
 & & & & & $\Xi_{c}^{+}+\Xi_{b}^{0}$ (8260.3) & $\Lambda K^{-} \pi^{+}\pi^{+}+?$ \\
 & & & & & & $\Sigma^{+} K^{-} \pi^{+}+?$ \\
 & & & & & & $\Xi^{0} \pi^{+}+?$ \\
 & & & & & & $\Xi^{-} \pi^{+} \pi^{+}+?$ \\
 & & & & & & $\Xi^{0} \pi^{+} \pi^{0}+?$ \\
 & & & & & & $\Xi^{0} \pi^{+} \pi^{+} \pi^{-}+?$ \\
\hline
\end{tabular}
\end{center}
\label{decay_dibaryon4}
\end{table*}%

\begin{table*}[p]
\caption{[$qq\,qQ\,QQ'$; $Q\!=\!c$, $Q'\!=\!b$ ($H_{bcc}$)] The possible decay modes of $H_{bcc}$-dibaryons with configuration $qq\,qQ\,QQ'$ ($Q\!=\!c$, $Q'\!=\!b$) against two baryon state $B$ and $B'$. The numbers in the brackets are invariant mass of the two baryons at rest frame.}
\begin{center}
\begin{tabular}{|c|c|c|c|c|c|c|}
\hline
flavor & config. of $H_{bcc}$ & $I$ & $I_{z}$ & $S$ & config. of $B+B'$ (MeV) & final states \\
\hline
21021 & $ud\,uc\,cb$ & 1/2 & 1/2 & 0 & $\Lambda_{b}^{0}+\Omega_{cc}^{+}$ (?) & $\Lambda_{c}^{+} \pi^{-}+?$ \\
\cline{6-7}
 & & & & & $\Lambda_{c}^{+}+\Xi_{cb}^{+}$ (?) & $p K^{-} \pi^{+} + ?$ \\
\hline
12021 & $ud\,dc\,cb$ & 1/2 & -1/2 & 0 & $\Lambda_{b}^{0}+\Omega_{cc}^{0}$ (?) & $\Lambda_{c}^{+}\pi^{-}+?$ \\
\cline{6-7}
 & & & & & $\Lambda_{c}^{+}+\Xi_{cb}^{0}$ (?) & $p K^{-} \pi^{+}+?$ \\
\hline
20121 & $us\,uc\,cb$ & 1 & 1 & -1 & $\Xi_{b}^{+}+\Omega_{cc}^{+}$ (?) & $?+?$ \\
\cline{6-7}
 & & & & & $\Lambda_{c}^{+}+\Xi_{cb}^{+}$ (?) & $\Lambda K^{-} \pi^{+}\pi^{+}+?$ \\
 & & & & & & $\Sigma^{+} K^{-} \pi^{+}+?$ \\
 & & & & & & $\Xi^{0} \pi^{+}+?$ \\
 & & & & & & $\Xi^{-} \pi^{+} \pi^{+}+?$ \\
 & & & & & & $\Xi^{0} \pi^{+} \pi^{0}+?$ \\
 & & & & & & $\Xi^{0} \pi^{+} \pi^{+} \pi^{-}+?$ \\
\hline
11121 & $\frac{1}{\sqrt{2}}(us\,dc + ds\,uc)\,cb$ & 1 & 0 & -1 & $\Lambda_{b}^{0}+\Omega_{cc}^{++}$ (?) & $\Lambda_{c}^{+}\pi^{-}+?$ \\
\cline{6-7}
 & & & & & $\Lambda_{c}^{+}+\Omega_{cb}^{0}$ (?) & $p K^{-} \pi^{+}+?$ \\
\hline
02121 & $ds\,dc\,cb$ & 1 & -1 & -1 & $\Xi_{b}^{0}+\Xi_{cc}^{+}$ (?) & $?+?$ \\
\cline{6-7}
 & & & & & $\Xi_{c}^{0}+\Xi_{cb}^{0}$ (?) & $p K^{-} K^{-} \pi^{+}+?$ \\
 & & & & & & $\Lambda K^{-} \pi^{+}\pi^{+}\pi^{-}+?$ \\
 & & & & & & $\Xi^{-}\pi^{+}+?$ \\
 & & & & & & $\Xi^{-}\pi^{+}\pi^{+}\pi^{-}+?$ \\
 & & & & & & $\Omega^{-}K^{+}+?$ \\
\hline
11121 & $ud\,sc\,cb$ & 0 & 0 & -1 & $\Lambda_{b}^{0}+\Omega_{cc}^{+}$ (?) & $\Lambda_{c}^{+}\pi^{-}+?$ \\
\cline{6-7}
 & & & & & $\Lambda_{c}^{+}+\Xi_{cb}^{+}$ (?) & $p K^{-} K^{-} \pi^{+}+?$ \\
\hline
\end{tabular}
\end{center}
\label{decay_dibaryon5}
\end{table*}%

\section{Summary}

Based on the consideration of the color-spin interaction between
diquarks, which describes well the mass splittings between
many hadrons and their spin flipped partners, we have systematically investigated possible stable multiquark configurations.  We have shown that the configurations becomes stable only when heavy quarks are included.
We have identified several new possibly stable multiquark configurations.
These are the $T^0_{cb}(ud \bar{c}\bar{b})$  tetraquark, the $\Theta_{bs}(udus \bar{b})$ pentaquark and the $H_c(udusuc)$ dibaryon, and so forth.  The representative decay modes for these states are $T_{cb}^{0}\rightarrow K^{+}\pi^{-}+K^{+}\pi^{-}$, $\Theta_{bs}\rightarrow \Lambda K^{+}\pi^{-}$ and
$H_c \rightarrow \Lambda+\Lambda_{c}^{+} \rightarrow p \pi^{-}+pK^{-}\pi^{+}\pi^{0}$.
If one observes these states, these will be the first explicit flavor multiquark states, and will be a valuable stepping stone to understanding QCD at larger quark number.

 The work of SHL was supported by
the Korea Research Foundation KRF-2006-C00011.


%

\end{document}